\documentclass[usenatbib, usegraphicx, useAMS]{mn2e}
\usepackage[colorlinks,urlcolor=blue,citecolor=blue,linkcolor=blue]{hyperref}
\usepackage{color}
\usepackage[T1]{fontenc}
\usepackage{aecompl}
\usepackage{breakurl}

\newcommand{\herschel}{\textit{Herschel}}
\newcommand{\swift}{\textit{Swift}}
\newcommand{\lx}{$L_{\rm X}$}
\newcommand{\mstar}{$M_{*}$}
\newcommand\aj{{AJ}}          \newcommand\araa{{ARA\&A}}          \newcommand\apj{{ApJ}}          \newcommand\apjl{{ApJL}}          \newcommand\apjs{{ApJS}}                    \newcommand\aap{{A\&A}}                              \newcommand\mnras{{MNRAS}}            \newcommand\nar{{New A Rev.}}  \newcommand\pasp{{PASP}}                    \newcommand\nat{{Nature}}          \newcommand\ssr{\ref@jnl{Space~Sci.~Rev.}}                    
\title[AGN Host Galaxy sSFR]{Decreased Specific Star Formation Rates in AGN Host Galaxies\thanks{{\it Herschel} is an ESA space observatory with science instruments provided by European-led Principal Investigator consortia and with important participation from NASA.}}
\author[T.~T. Shimizu et al.]{T. Taro Shimizu$^1$\thanks{Email: tshimizu@astro.umd.edu}, Richard F. Mushotzky$^1$,  Marcio Mel\'endez$^1$,  Michael Koss$^2$,
\newauthor and David J. Rosario$^3$ \\
$^1$Department of Astronomy, University of Maryland, College Park, MD 20742, USA \\
$^2$Institute for Astronomy, Department of Physics, ETH Zurich, Wolfgang-Pauli-Strausse 27, CH-8093 Zurich, Switzerland \\
$^3$Max-Planck-Institut f{\"u}r Extraterrestrische Physik (MPE), Postfach 1312, D-85741 Garching, Germany}

\begin{document}
\maketitle
\begin{abstract}
We investigate the location of an ultra-hard X-ray selected sample of AGN from the \textit{Swift} Burst Alert Telescope (BAT) catalog with respect to the main sequence (MS) of star-forming galaxies using \herschel-based measurements of the SFR and \mstar's from Sloan Digital Sky Survey (SDSS) photometry where the AGN contribution has been carefully removed.  We construct the MS with galaxies from the \herschel{} Reference Survey and \herschel{} Stripe 82 Survey using the exact same methods to measure the SFR and \mstar{} as the \textit{Swift}/BAT AGN. We find a large fraction of the \textit{Swift}/BAT AGN lie below the MS indicating decreased specific SFR (sSFR) compared to non-AGN galaxies. The \textit{Swift}/BAT AGN are then compared to a high-mass galaxy sample (COLD GASS), where we find a similarity between the AGN in COLD GASS and the \textit{Swift}/BAT AGN. Both samples of AGN lie firmly between star-forming galaxies on the MS and quiescent galaxies far below the MS. However, we find no relationship between the X-ray luminosity and distance from the MS. While the morphological distribution of the BAT AGN is more similar to star-forming galaxies, the sSFR of each morphology is more similar to the COLD GASS AGN. The merger fraction in the BAT AGN is much higher than the COLD GASS AGN and star-forming galaxies and is related to distance from the MS. These results support a model in which bright AGN tend to be in high mass star-forming galaxies in the process of quenching which eventually starves the supermassive black hole itself.
\end{abstract}

\begin{keywords}
galaxies: active -- galaxies: nuclei -- galaxies: Seyfert -- stars: formation -- infrared: galaxies
\end{keywords}

\section{Introduction}\label{intro}
The link between supermassive black holes (SMBH) and their host galaxies has been evident for many years through the study of correlations between large scale host galaxy properties and SMBH mass. Tight correlations were found with bulge stellar velocity dispersion \citep{Ferrarese:2000gf, Gebhardt:2000xy, Gultekin:2009ul}, bulge luminosity \citep[e.g.][]{Magorrian:1998qv}, and bulge mass \citep{Kormendy:1995mz, Marconi:2003ve, Haring:2004ly} suggesting a coevolution of the SMBH with the host galaxy. The question that remains however is how two seemingly disjoined objects can influence each other over an extremely large range of physical scales. Using a simple estimate of the sphere of influence for a SMBH ($r_{sph} = GM/\sigma^2$; \citet{Peebles:1972nr}) and a typical SMBH mass of $10^8 M_{\sun}$ and stellar velocity dispersion, $\sigma=200$ km s$^{-1}$, $r_{sph} \sim 10$ pc whereas the size of the bulge is roughly several kpc \citep[e.g.][]{Simard:2011sf}. Therefore any influence from the SMBH must be able to extend over 3 orders of magnitude in physical scale. 

Active galactic nuclei (AGN), the phase where the SMBH is vigorously accreting material and growing, are thought to supply the necessary energy to influence the galaxy on large scales \citep[e.g.][]{Silk:1998qf}. This leads to a feedback cycle wherein the galaxy supplies cold gas that ignites the AGN and fuels star formation, and the AGN then returns energy and/or momentum to the galaxy that shuts off both accretion and star formation. The explicit feedback mechanism that runs this cycle is currently not well understood but is thought to be either large scale outflows  \citep{Kaviraj:2011kl, Cimatti:2013gd, Veilleux:2013qq, Harrison:2014xe} or radio jets \cite{Best:2007vn, Dubois:2013vn}\citep[for a review see][]{Fabian:2012dp}. Theoretical simulations have also shown that adding AGN feedback reproduces well the observed mass and luminosity functions while an absence produces too many blue, high mass galaxies \citep[e.g.][]{Croton:2006kx}.

Early evidence for the quenching of star formation due to AGN came from studying the colors and stellar masses of large samples of both non-AGN and AGN host galaxies. Whereas the non-AGN sample clearly separates into two populations on a color-magnitude or color-mass diagram, one with red colors (i.e. red sequence thought to be quiescent galaxies) and one with blue (i.e. blue cloud thought to be strongly star-forming galaxies) \citep{Strateva:2001tw}, AGN host galaxies were found to be concentrated between them, displaying ``green'' colors \citep[e.g][]{Nandra:2007rz, Silverman:2008zl, Hickox:2009ys}. Initially, this was interpreted as AGN preferentially occurring in galaxies that have had their star formation recently quenched \citep{Salim:2007rm, Martin:2007yg, Schawinski:2009zr} and are in transition from the blue cloud to the red sequence. Optical colors, however, can be imprecise tracers of star formation, especially in the presence of strong intrinsic dust absorption \citep[e.g.][]{Cardamone:2010uq} that obscures recent star formation and causes a reddening of the colors that is not due to a reduction of star-formation. Green colors can also just be an indication of a mixture of old and new stellar populations and not necessarily a ``transition'' between the two population. Further, recent studies that mass-match non-AGN galaxy samples to AGN host galaxies reveal that the difference in optical colors virtually disappear \citep{Silverman:2009lq, Pierce:2010qy, Rosario:2013qf, Rosario:2013ye}

Far-infrared (FIR) emission ($\lambda > 40\micron$) is essentially immune to reddening effects while also being a direct tracer of recent star formation. Dust in the galaxy is heated by UV photons from recently formed OB stars, that then reemit in the mid-far infrared regime \citep{Draine:2003gd} creating a strong correlation between the FIR luminosity and SFR of a galaxy \citep{Kennicutt:1998kx, Kennicutt:2012it}. Furthermore, AGN are not thought to strongly affect the FIR \cite[e.g.][]{Netzer:2007ve}, and thus the FIR is more robust compared to other SFR indicators such as UV continuum and H$\alpha$ line emission that are mainly used in non-AGN galaxy studies.

The \herschel{} \textit{Space Observatory} \citep{Pilbratt:2010rz} opened a window into the FIR universe with the unprecedented sensitivity of both the Photodetector Array Camera and Spectrometer (PACS; \citet{Poglitsch:2010fp}) and Spectral and Photometric Imaging Receiver (SPIRE;  \citet{Griffin:2010sf}) instrument extending the broad band spectral energy distributions out to 500 \micron{} and allowing an accurate estimate of the FIR luminosity of more objects than allowed by previous FIR telescopes (i.e. \textit{Infrared Astronomical Satellite (IRAS)}; \citet{Neugebauer:1984fp}). In this study we utilize \herschel{} to measure the SFRs of a large and relatively unbiased sample of AGN and compare their location on the SFR-stellar mass (\mstar) diagram with that of the general star-forming galaxy population, which forms a ``main sequence''.

The main sequence is the observed tight correlation between the stellar mass and SFR of a normal star-forming galaxy and has been confirmed in depth by many studies \citep[e.g.][]{Brinchmann:2004bf, Noeske:2007fr, Elbaz:2007ud, Rodighiero:2010ef, Elbaz:2011qe, Whitaker:2012uq, Magnelli:2014sf}. This correlation seems to exist up to at least $z\sim2$ \citep[e.g.][]{Elbaz:2011qe, Whitaker:2012uq} and possibly all the way to $z\sim4$ \citep{Bouwens:2012fv, Heinis:2014yg}, with only the normalization changing as a function of redshift, shifting to higher SFRs at earlier epochs. This discovery has changed theories of galaxy evolution from one that is merger-driven to one that is driven more by internal secular processes. \citet{Elbaz:2011qe} showed that galaxies that live above the main sequence are much more compact with a higher SFR surface density, indicative of a major merger. Main sequence galaxies, though, have disk-like morphologies inconsistent with a recent merger \citep{Wuyts:2011lh} that suggests star formation is triggered by internal processes such as disk instabilities. 

In the past, the question of where AGN fit into the picture was unclear because most main sequence studies purposely excluded AGN due to its messy contribution to SFR indicators. However, with \herschel{}, more accurate estimates of the SFRs for AGN host galaxies can be calculated as well as better detection rates. For example, both \citet{Mullaney:2012fk} and \citet{Rosario:2013ye}, using deep observations of large fields examined the star-forming properties of X-ray selected AGN. Both came to the conclusion that AGN primarily reside in main sequence galaxies, calling into question the long-held idea that AGN host galaxies are in the process of quenching star formation. If most AGN are in main sequence galaxies, this could indicate that moderate luminosity AGN are simply coincidental with a large cold gas reservoir that also fuels star formation \citep{Vito:2014eu}.

In this Paper, we seek to fully investigate the location of AGN on the SFR-\mstar{} plane using \herschel{} observations of \swift/BAT selected AGN. These are ultra-hard X-ray confirmed AGN at low redshift using a selection method that is unbiased with respect to both obscuration and host galaxy contamination. We calculate SFRs using \herschel{} photometry and a simple model to disentangle star formation and AGN contributions and combine them with estimates of the stellar mass using AGN subtracted SDSS photometry from \citet{Koss:2011vn}. The SFRs and \mstar's are then compared to a local normal star-forming galaxy sample to define the main sequence as well as a sample of galaxies purely selected on stellar mass. Finally we discuss the implications of our results and compare them to previous studies. Throughout this paper we use a $\Lambda$CDM cosmology with $H_{0} = 70 \rm{km\,s^{-1}\,Mpc^{2}}$, $\Omega_{m} = 0.3$, and $\Omega_{\Lambda} = 0.7$. Luminosity distances were calculated using this cosmology along with redshifts taken mainly from the \textit{NASA/IPAC} Extragalactic Database (NED)\footnote{\url{http://ned.ipac.caltech.edu/}}, except for those objects with $z < 0.01$ where we used measured distances from the Extragalactic Distance Database\footnote{\url{http://edd.ifa.hawaii.edu/}}. We either use or correct for a Chabrier or Kroupa initial mass function (IMF) for all star formation rate calculations.

\section{Samples and Observations}
\subsection{\textit{Swift}/BAT AGN}
Our parent sample of AGN was drawn from the 58 month \swift{} Burst Alert Telescope (BAT) \citep{Gehrels:2004qf,Barthelmy:2005ul} catalog \citep{Baumgartner:2013fq} with a redshift cutoff of $z<0.05$, totaling 313 AGN (149 Seyfert 1-1.5s, 157 Seyfert 1.8-2s, 6 LINERs, and 1 unidentified AGN). The catalog is the result of continuous monitoring by \swift/BAT of the entire sky in the 14--195 keV energy range. These high energies allow for an unambiguous detection of AGN with little to no contamination from the host galaxy and significantly reduced selection effects due to obscuration. 

All 313 AGN were observed by the \herschel{} \textit{Space Observatory} with 291 part of our program (PI: R. Mushotzky, PID: OT1\_rmushotz\_1) and the remaining 22 obtained from other programs publicly available on the \herschel{} Science Archive. The sample was imaged by both the PACS  and SPIRE instruments providing, for the first time, sensitive FIR photometry from 70--500 \micron{} for a large, ultra-hard X-ray selected sample of AGN. Detailed descriptions of the reduction and analysis of the PACS and SPIRE images are given in \citet{Melendez:2014yu} and Shimizu et al (2015, in preparation), but we provide a brief summary here.

PACS and SPIRE together imaged the sample in 5 broad bands: 70 and 160 \micron{} (PACS) and 250, 350, and 500 \micron{} (SPIRE). Level 0 (raw) data were reduced to Level 1 using the standard pipeline provided by the \herschel{} Interactive Processing Environment \citep{Ott:2010rm} v8.0. Maps were produced from the Level 1 data using \textit{Scanamorphos} \citep{Roussel:2013gf} v19.0, a software package that takes advantage of the redundancy inherent in the scanning procedure of \herschel{} to remove both thermal and non-thermal low-frequency noise. Circular and elliptical apertures with radii chosen visually to encompass the entirety of the FIR emission were then used to extract the photometry for each waveband. 1$\sigma$ errors for the photometry were determined using a combination of the pixel-by-pixel errors in the aperture, an estimate of the root-mean-square of the background, and calibration uncertainty. 

In addition to \herschel{} observations, \citet{Koss:2011vn} analyzed optical images of 185 BAT AGN from the \textit{Sloan Digital Sky Survey} (SDSS) and Kitt Peak National Observatory. Using GALFIT \citep{Peng:2002dq}, they were able to accurately measure the host galaxy light by subtracting out the central point source due to the AGN. Reliable stellar masses for the BAT AGN host galaxies were then estimated using standard stellar population models. Because the \citet{Koss:2011vn} BAT sample was chosen from the 22 month catalog \citep{Tueller:2010ul}, 45/185 were not observed with \herschel, reducing the sample to 140 AGN. Furthermore, \citet{Koss:2011vn} flagged 18/140 objects for incomplete PSF-subtraction from the \textit{griz} images, so we choose not to include these sources resulting in a final sample of 122 AGN, including 46 Sy 1s, 72 Sy 2s, and 4 LINERs, where we define a Sy 1 as Sy 1-1.5 and Sy 2 as Sy 1.8-2. The reason for the discrepancy between the number of Sy 1s and Sy 2s is that all 18 of the objects that were flagged for incomplete PSF-subtraction are Sy 1s.

\subsection{\herschel{} Reference Survey}
To form the main sequence, we need a large and complete sample of star-forming galaxies that do not host an AGN but have been observed at the same wavelengths allowing for a consistent determination of both the SFR and stellar mass. For these reasons we chose the \textit{Herschel Reference Survey} (HRS; \citet{Boselli:2010fj}), a guaranteed time \herschel{} key project that imaged 323 K-band selected galaxies from 100--500 \micron{}. The HRS spans all morphological types and was volume limited to contain galaxies between 15 and 25 Mpc away. Even though our sample stretches out to $z=0.05$ ($\sim$200 Mpc), the HRS represents the best sample of local star-forming galaxies to compare with given that both have been observed by \herschel{} as well as other telescopes including the \textit{Wide-Field Infrared Survey Explorer} \citep[WISE]{Wright:2010fk}, the \textit{Galaxy Evolution Explorer} \citep[GALEX][]{Martin:2005uq}, and \textit{SDSS}.

The HRS PACS and SPIRE images were analyzed in \citet{Cortese:2014qq} and \citet{Ciesla:2012lq} producing photometry at 100, 160, 250, 350, and 500 \micron. We applied the same corrections to the SPIRE photometry as described in \citet{Cortese:2014qq} due to changes in the SPIRE calibration and beam size. The corrections reduce the SPIRE flux densities by 7, 6, and 9 per cent at 250, 350, and 500 \micron{} respectively. 

A subset of the HRS are galaxies within the Virgo Cluster and have been affected by the dense environment through the stripping of their gas \citep{Boselli:2006qf}. The dust content of these galaxies has also been shown to be affected by the environment \citep{Cortese:2010jk, Cortese:2012xy}. Therefore, following \citet{Ciesla:2014uo}, we restrict the HRS sample to only those galaxies that aren't ``HI-deficient'' as defined in \citet{Boselli:2012qv} which reduces the sample to 146 galaxies.

\subsection{COLD GASS}\label{sec:cold_gass}
While the HRS represents a sample that was observed using the same telescopes, it is limited in its range of stellar mass especially above $10^{10} M_{\sun}$. For a complete comparison to the BAT AGN, we supplemented HRS with the \textit{CO Legacy Database for GASS} \citep[COLD GASS;][]{Saintonge:2011lr}, a 366 galaxy subsample of the \textit{GALEX Arecibo SDSS Survey} \citep[GASS;][]{Catinella:2010fj}. The GASS sample consists of $\sim$1000 galaxies randomly selected such that every galaxy lies within the footprint of the SDSS spectroscopic survey, the ALFALFA survey, and the \textit{GALEX} Medium Imaging Survey. The galaxies also were selected to have a redshift range $0.025<z<0.05$ and a stellar mass between $10^{10}$ and $10^{11.5}$, both of which match very well to the BAT AGN. 366 galaxies were randomly selected from GASS to form the COLD GASS sample and have deep CO(1--0) imaging with IRAM. COLD GASS then represents a completely unbiased sample of galaxies above $10^{10} M_{\sun}$.

As described in \citet{Catinella:2010fj} and \citet{Saintonge:2011lr}, both GASS and COLD GASS were selected to have a uniform $\log$\mstar{} distribution. The stellar mass distribution that is observed, however, is more heavily weighted towards lower mass galaxies (see Figure 1 of \citet{Saintonge:2011lr}). Therefore, \citet{Saintonge:2011lr} constructed 50 representative subsamples of COLD GASS that matches the observed \mstar{} distribution. Each subsample contains between 200-260 galaxies and can be used to test the robustness of any relation that might be observed. Throughout this Paper, we explicitly note whether the full COLD GASS sample, an average of the representative subsamples, or a single representative subsample is being used.
 
Because COLD GASS was selected in an unbiased way, the sample contains a mixture of galaxy types from star-forming to quiescent to AGN. To determine the type of each galaxy in COLD GASS, we cross-matched the sample with the MPA-JHU SDSS DR7 spectroscopic catalog.\footnote{\url{http://www.mpa-garching.mpg.de/SDSS/DR7/} and \url{http://home.strw.leidenuniv.nl/~jarle/SDSS/}} In this catalog, galaxies that have SDSS optical spectra were classified according to their location on the standard BPT diagram \citep{Baldwin:1981yq}. \citet{Brinchmann:2004bf} describes the details of the classification\footnote{We note that \citet{Saintonge:2012uq} used a more conservative method to classify AGN in the COLD GASS. A galaxy was considered an AGN if $\log$[NII/H$\alpha$]$>-0.22$ and $\log$[OIII/H$\beta$]$>0.48$. This led to a much lower fraction of AGN in their analysis (6 per cent) compared to this work (30 per cent).}. Galaxies were separated into 6 groups: Star-forming, Low S/N Star-forming, AGN, Composite, LINER, and Quiescent. For the purposes of comparison with the BAT AGN, we combined the Star-forming and Low S/N Star-forming groups into a single Star-forming group and the AGN and Composite groups into a single AGN+Composite group. We combine the AGN and Composite group because after cross-matching the BAT AGN with the same SDSS DR7 sample, we find that almost all are classified as an AGN or Composite galaxy. The COLD GASS AGN+Composite subsample then represents an ideal optically selected sample of AGN to compare with the BAT AGN, while the Star-forming and Quiescent sample represent ideal non-AGN samples. We denote the different subsamples in the following way: Star-forming as CGS, AGN+Composite as CGA, LINERs as CGL, and Quiescent as CGQ.  

\subsection{\herschel{} Stripe 82 Survey}
With the HRS and COLD GASS samples we have one that matches the photometry available to the BAT AGN (HRS) and one that more closely matches the physical properties of the BAT AGN host galaxies (COLD GASS). HRS lacks galaxies at the high stellar mass of the BAT AGN whereas COLD GASS lacks \herschel{} and \textit{WISE} photometry to allow for a consistent comparison between SFRs. Therefore, we use a third comparison sample to the BAT AGN, the \herschel{} Stripe 82 survey \citep[HerS;][]{Viero:2014jk}that bridges the gap between HRS and COLD GASS.

HerS is a 79 deg$^{2}$ survey of SDSS Stripe 82 with \herschel/SPIRE at 250, 350, and 500 \micron. \citet{Viero:2014jk} produced the catalog of HerS sources while Rosario et al (2015, in preparation) calculated SFRs based on FIR luminosities. Given its small area, the volume covered at low redshift is smaller than that of both the COLDGASS and BAT samples which means HerS is rather incomplete at high stellar masses ($>10.5 M_{\sun}$). However it is still better than the HRS and has \herschel{} photometry available, albeit only from the SPIRE instrument. 

We selected our HerS sample from Rosario et al (2015, in preparation), who matched HerS sources from \citet{Viero:2014jk} with the SDSS MPA-JHU DR7 catalog. Because the HerS catalog assumes all galaxies are point sources, Rosario et al (2015, in preparation) limited their sample to $z > 0.02.$ We further limit the sample to $z < 0.08$ to match the BAT AGN while also pushing out to a slightly larger volume to populate the high mass end better without a significantly affecting our results. After cutting sources which do not have a measured stellar mass as well as sources with low S/N ($<3$) emission lines (only 5 objects) this results in a final HerS sample of 517 objects. Due to the combined magnitude cut from the SDSS and the limited sensitivity of SPIRE from which the HerS catalog was built, Malmquist bias can be a problem. However, because of the relatively low redshift nature of our sample we do not expect it to largely bias our results.

The HerS sample was also split into some of the same classifications as COLDGASS using the BPT diagram. HerS contains both a star-forming and AGN population and will be designated as HerS SF and HerS AGN. The HerS AGN, just as CGA, is a combination of the AGN and Composite classifications. Within HerS, there is also an Uncertain classification which indicates a galaxy that is missing or has upper limits for at least one of the 4 key lines needed to classify it using the BPT diagram.

\section{Star Formation Rate Estimation}\label{sec:sfr_estimates}
The unique \herschel{} data provides a means for determining accurate star formation rates (SFR). IR emission has long been used as a calibrator for star formation \citep[see][for a review]{Kennicutt:1998kx}, because it probes the dust population that reprocesses the UV emission from young stars \citep[e.g.][]{Draine:2003gd}. The specific wavelength range of \herschel{} covers the bulk of the IR emission from dust including the characteristic FIR bump typically seen in star-forming galaxies \citep[e.g.][]{Dale:2007fk} allowing for precise measures of the total IR luminosity and thus the star formation rate, especially for AGN host galaxies where many of the often used SFR indicators (e.g. H$\alpha$, UV continuum) can be substantially contaminated by AGN-related emission.

Many SED-fitting packages exist in the literature ranging from template based models to full dust radiative transfer models. However, given the low number of data points for our SEDs (at most seven), we chose to fit our SEDs with the model described in \citet{Casey:2012jl}, which is a combination of an exponentially cutoff mid-infrared (MIR) power law and a single temperature greybody. Details and results of the SED fitting for the BAT AGN will be given in a forthcoming paper (Shimizu et al 2015, in preparation), however we provide example fits to both the BAT AGN and HRS in Appendix~\ref{sec:appendix} as well as a brief overview of the fitting procedure. In Section~\ref{subsec:sel_effects_mod_depend} we discuss extensively and test whether this model introduces systematic biases especially related to the decomposition of the SED. 

The \herschel{} photometry constrains the greybody component, but we need additional shorter wavelength data to constrain the power law component. Therefore, we cross-correlated our sample with the \textit{Wide-field Infrared Survey Explorer} \citep[\textit{WISE}]{Wright:2010fk} All-Sky Release Catalog on the IRSA website.\footnote{\url{http://irsa.ipac.caltech.edu/Missions/wise.html}} We only use the 12 (W3) and 22 (W4) \micron{} photometry to avoid contamination by the stellar population in the host galaxy. \textit{WISE} photometry for the HRS were taken from \citet{Ciesla:2014uo}.

For consistency, we fit both the BAT AGN and HRS galaxies using the \citet{Casey:2012jl} model even though the HRS galaxies do not host an AGN or are classified as ULIRGs. In this way, we can account for a portion of MIR emission that is due to normal star formation rather than AGN heated dust. We fit all BAT AGN and HRS galaxies with at least four detected photometric points. This restriction ensures quality photometry for each galaxy and removes only a further 9 and 11 galaxies from the BAT AGN and HRS sample respectively for final sample sizes of 113 and 135. We determine three luminosities for each galaxy: a total IR luminosity ($L_{\rm TIR}$), a MIR power law luminosity ($L_{\rm MIR}$) and a greybody luminosity ($L_{\rm Grey}$). Each one was calculated by integrating the best-fitting model from 8--1000 \micron{}. $L_{\rm TIR}$ is the luminosity from integrating over the total model while $L_{\rm MIR}$ is from only integrating the MIR power law component, and $L_{\rm Grey}$ is from the greybody component. 

Star formation rates are then calculated using one of these IR luminosities and the equation from \citet{Murphy:2011rt}.
\begin{equation}\label{ir_sfr}
\rm{SFR}_{\rm IR} =   \frac{L_{\rm IR}\,\,[\rm{erg}\,\,\rm{s}^{-1}]}{2.57\times10^{43}}
\end{equation}
\noindent For the HRS galaxies $L_{\rm IR} = L_{\rm TIR}$ since there is no AGN to contribute to the IR emission. For the BAT AGN, however, we use $L_{\rm IR} = 4/3L_{\rm Grey}$. The 4/3 is a correction factor to account for MIR emission from star formation. To determine it, we calculated the average ratio of $L_{\rm MIR}$/$L_{\rm Grey}$ for the HRS sample, which contain no AGN, and found it to be narrowly distributed around 1/3. This means that only using $L_{\rm Grey}$ to determine the SFR underestimates it by 1/3 so we need to multiply $L_{\rm Grey}$ by 4/3 as a correction.

While FIR emission probes dust obscured star formation, the UV continuum is a measure of unobscured star formation by tracing the direct light from young massive stars. Hence, a complete census of star formation can be found by combining measurements from both wavebands. AGN are strong emitters in the UV though, so only Seyfert 2 galaxies will have reliable UV photometry to combine with the FIR for a SFR. Using \textit{GALEX} far-UV data from the GCAT catalog \citep{Bianchi:2014qy} for the BAT AGN Seyfert 2's and the same from \citet{Boselli:2013kx} for the HRS, we calculated dust-corrected UV SFRs. We found that using UV SFRs for both the HRS and BAT AGN Seyfert 2s had no effect on the results of this Paper. Thus, we choose to use the FIR only SFRs to allow a larger BAT AGN sample (Seyfert 1's and 2's). 

The COLD GASS sample unfortunately was not observed with \herschel{} and does not allow for the same calculation of the SFR. We use the SFRs provided in \citet{Saintonge:2011fk} which were calculated by fitting SDSS and \textit{GALEX} photometry to \citet{Bruzual:2003lr} models. \citet{Saintonge:2011fk} compared these SFRs to those inferred from combined \textit{GALEX} FUV and \textit{Spitzer} 70 \micron{} data finding a strong correlation with only a scatter of 0.22 dex and indicating FIR-based SFRs are consistent with optical-UV ones. We recognize that the \citet{Saintonge:2011fk} comparison however did not use any \herschel{} photometry, but three calibration analyses\footnote{\url{http://herschel.esac.esa.int/twiki/pub/Public/PacsCalibrationWeb/ExtSrcPhotom.pdf}}\footnote{\url{http://herschel.esac.esa.int/twiki/bin/viewfile/Public/PacsCalibrationWeb?rev=1;filename=PICC-NHSC-TN-029.pdf}}\footnote{\url{https://nhscsci.ipac.caltech.edu/pacs/docs/Photometer/PICC-NHSC-TR-034.pdf }} show that \textit{Spitzer} and \herschel{} produce consistent fluxes. Further \citet{Dominguez-Sanchez:2014dn} recently showed that SFRs calculated from FIR SED fitting are consistent (with large scatter especially for pure AGN and quiescent galaxies) with the SFRs inferred from the MPA-JHU SDSS spectral analysis. Based on all these indirect tests, we are confident that the COLD GASS SFRs are consistent with the \herschel{} based ones we use for the HRS and BAT AGN in a way that does not effect the results of this Paper.

For the HerS sample, the SFRs were estimated by fitting the SPIRE 250 \micron{} and \textit{WISE} 22 \micron{} fluxes to the \citet{Dale:2002ty} (DH02) templates. Each of the 64 templates represents a different value of $\alpha$ where $\alpha$ is the power law index for the distribution of dust mass over heating intensity in a galaxy. $\chi^{2}$ minimization was used to scale each of the 64 templates to the observed fluxes in the HerS sample, then the template with the lowest $\chi^{2}$ was chosen as the best fit. The best-fit template was integrated between 8--1000 \micron{} to calculate $L_{\rm IR}$ and converted to a SFR using the SFR--$L_{IR}$ relation from \citet{Kennicutt:1998kx} adjusted for a Chabrier IMF (lowered by a factor of 1.7). In the absence of \textit{WISE} 22 \micron{} photometry, $\alpha$ was fixed at -2.0.

This is a slightly different method than the one used for the HRS and BAT AGN. To test for systematics we fit the HRS galaxies with 250 \micron{} and 22 \micron{} detections using the DH02 method and compared the SFRs. A linear fit to the two SFRs reveals a slope of 1.0 and an offset of 0.11 dex with the SFRs determined from the \citet{Casey:2012jl} and Equation~\ref{ir_sfr} higher than the ones from DH02 and the relation from \citet{Kennicutt:1998kx}. Therefore we adjust all of the HerS SFRs by adding a constant of 0.11 dex.

\section{Stellar Mass Estimates}
SFRs are only one-half of the main sequence; stellar masses are also needed for the galaxies. \citet{Cortese:2012fj} calculated the stellar masses of the HRS using the relation from \cite{Zibetti:2009jf}:
\begin{equation}\label{eq:stellar_mass}
\log(M_{*}/L_{i}) = -0.963 + 1.032(g-i)
\end{equation}
\noindent where $M_{*}$ is the stellar mass and $L_{i}$ is the $i$-band luminosity, both in solar units. To be consistent we also used this equation for the BAT AGN. The $g-i$ color was calculated using the PSF-subtracted photometry from \citet{Koss:2011vn}. The stellar masses determined here correlate very well with the stellar masses from \citet{Koss:2011vn} with a Pearson correlation coefficient, $r_{\rm P} = 0.85$ indicating a highly linear relationship. However the \citet{Koss:2011vn} values are systematically larger by a factor of $\sim2$\footnote{This is most likely due to the different stellar population models used in \cite{Zibetti:2009jf}. They used the 2007 version of the \citet{Bruzual:2003lr} models that included a new presciption for thermally pulsing AGB stars. This decreases the stellar mass by a factor of 2 especially for star-forming galaxies. This has no effect on our results as long as all samples are on the same stellar mass scale.} Since the goal of this Paper is to compare AGN with normal star-forming galaxies and not absolute measures of stellar mass and SFR, we choose to use Equation~\ref{eq:stellar_mass} for the BAT AGN stellar masses. We performed the same calculation for the COLD GASS galaxies and compared these \mstar{} with those provided in \citet{Saintonge:2011fk}. We find the same strong correlation and the same systematic offset as the BAT AGN stellar masses so we choose to also use the stellar masses calculated in this Paper for COLD GASS as well. The HerS stellar masses are taken from the MPA-JHU database which used the same method as both \citet{Koss:2011vn} and \citet{Saintonge:2011fk}. However, we do not have $g$ or $i$ photometry for HerS so we apply a factor of 2 correction to them to match the stellar mass scale from Equation~\ref{eq:stellar_mass}.

\section{Results}
\subsection{Location of AGN in SFR-$M_{*}$ Plane}\label{subsec:agn_ms_loc}
One key issue in this analysis is which main sequence to use. Many authors have published main sequence relations \citep[for a nice compilation see:][]{Speagle:2014rc}, however each relation was determined differently using different stellar mass estimates, SFR indicators, and redshift ranges. This has resulted in a large spread of values for both the slope and normalization of the main sequence, especially in the local universe. Therefore, we choose to calculate our own main sequence relation using only the HRS galaxies and the HerS star-forming galaxies since both the stellar masses and SFRs were calculated with comparable methods as for the BAT AGN. We use a linear bisector \citep{Isobe:1990fk} to fit the HRS+HerS data, resulting in the following equation:

\begin{equation}\label{eq:ms_hrs_ir}
\log(\rm SFR_{\rm IR}) = 1.01 \log(M_{*}) - 9.87
\end{equation}
\noindent The scatter of the relation is 0.36 dex, similar to the scatter seen in other studies \citep[e.g][]{Noeske:2007fr, Peng:2010wd}. The slope and the normalization are slightly steeper and smaller respectively than that found in \citet{Peng:2010wd}, who analyzed the SFRs and stellar masses of the entire local ($0.02<z<0.085$) SDSS population. The slope is much steeper than the $z=0$ slope predicted using the \citet{Speagle:2014rc} relation (0.5). This is possibly due to the addition of lower mass objects from HRS as well differences in the measurement of the SFR and \mstar. Further, due to the large scatter in MS relations measured at low redshift, \citet{Speagle:2014rc} specifically did not include low redshift studies in formulating their redshft-dependent MS relation. This is the reason we set out to formulate our own main sequence relation that uses a well-defined star-forming galaxy sample and measures of the SFR and stellar mass that are consistent between the non-AGN and AGN host galaxy samples.
\begin{figure*}
\centering
\includegraphics[height=0.9\textheight]{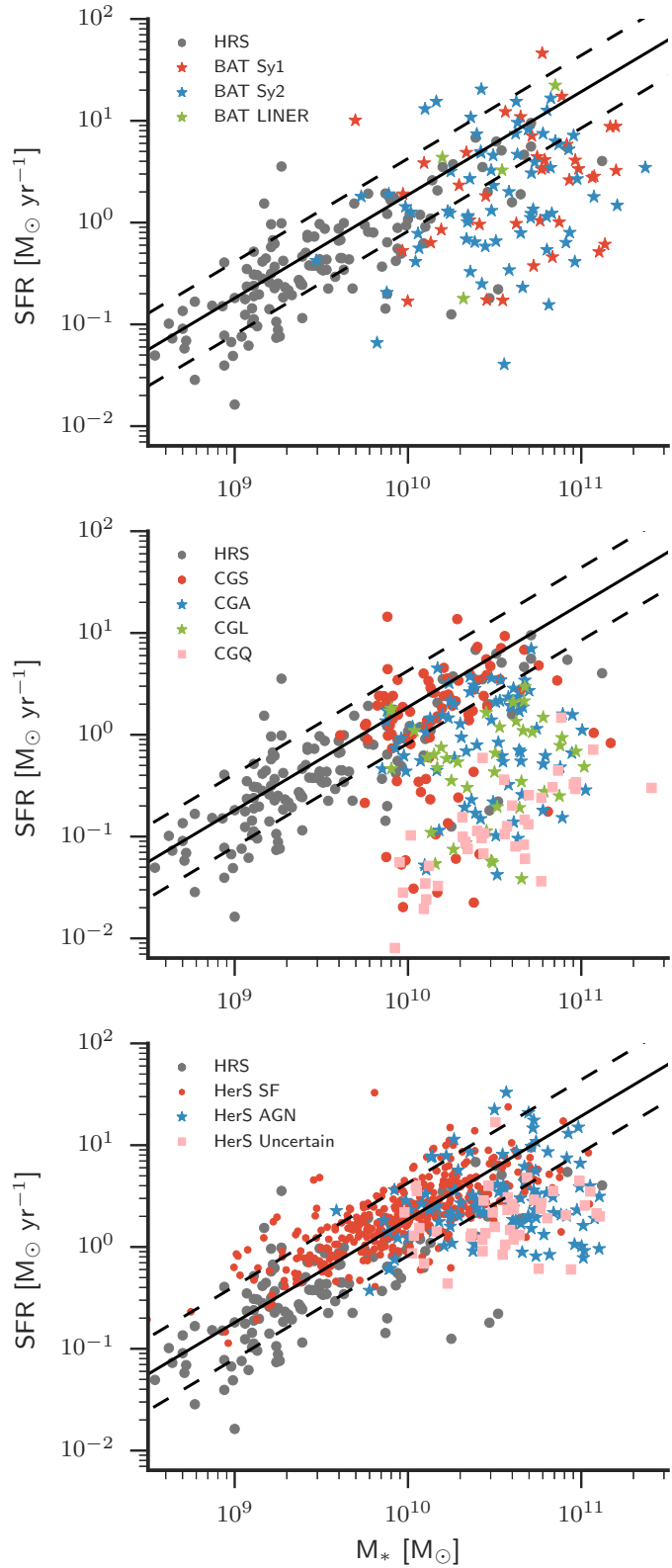}
\caption{The relationship between SFR and $M_{*}$ for the HRS (black dots), BAT AGN (\textit{top:} colored stars), a representative subsample of COLD GASS (\textit{middle:} colored symbols), and the HerS sample (\textit{bottom:} colored symbols). The solid line represents the main sequence relationship calculated using the IR-based HRS+HerS SFRs (Equation~\ref{eq:ms_hrs_ir}). The dashed lines are the measured $1\sigma$ (0.32 dex) scatter in the relationship.\label{fig:ms_plot}}
\end{figure*}
\begin{table*}
\begin{minipage}{6.0in}
	\caption{Main Sequence Locations for BAT AGN and COLD GASS}
	\label{tab:percentages}
	\begin{tabular}{@{}lcccccc}
		\hline
		
		Sample & Total &Above MS & Inside MS & $1-2\sigma$ Below & $2-3\sigma$ Below & $> 3\sigma$ Below \\
		(1) & (2) & (3) & (4) & (5) & (6) & (7) \\
		\hline
		\textbf{BAT AGN} \\
		All & ... & 0.05 & 0.28 & 0.28 & 0.18 & 0.20 \\
		Sy 1 & 0.35 & 0.05 & 0.20 & 0.35 & 0.20 & 0.20 \\
		Sy 2 & 0.61 & 0.06 & 0.30 & 0.26 & 0.17 & 0.20 \\
		LINER & 0.04 & 0.00 & 0.75  & 0.00 & 0.00 & 0.25 \\
		\\
		\textbf{COLD GASS (CG)}\\
		All & ... & 0.02 & 0.33 & 0.15 & 0.12 & 0.39 \\
		Star-Forming (CGS) & 0.37 & 0.04 & 0.62 & 0.12 & 0.05 & 0.16 \\
		AGN+Comps (CGA) & 0.30 & 0.02 & 0.28 & 0.23 & 0.15 & 0.34 \\
		LINER (CGL) & 0.17 & 0.00 & 0.07 & 0.18 & 0.25 & 0.50 \\
		Quiescent & 0.15 & 0.00 & 0.00 & 0.00 & 0.03 & 0.97 \\
		\\
		\textbf{\herschel{} Stripe 82 (HerS)}\\
		All & ... & 0.12 & 0.67 & 0.14 & 0.05 & 0.02 \\
		Star-Forming & 0.72 & 0.14 & 0.78 & 0.07 & 0.01 & 0.00 \\
		AGN+Comps & 0.19 & 0.08 & 0.48 & 0.25 & 0.10 & 0.09 \\		
		Uncertain & 0.08 & 0.02 & 0.21 & 0.42 & 0.28 & 0.07\\ 
		\hline
	\end{tabular}
	
	\medskip
	\textbf{Notes} (1) Sample and subsample names for the BAT AGN and COLD GASS. (2) Fraction of the total sample (i.e. BAT AGN and COLD GASS) that each subsample occupies. (3) Fraction of sample that is above the main sequence (MS) ($\Delta\log\rm{SFR} > 1\sigma$; $\sigma=0.32$ dex). (4) Fraction of sample that is inside the MS ($1\sigma > \Delta\log\rm{SFR} >  -1\sigma$). (5) Fraction of sample that is between 1$\sigma$ and 2$\sigma$ below the MS ($-1\sigma > \Delta\log\rm{SFR} >  -2\sigma$). (6) Fraction of sample that is between 2$\sigma$ and 3$\sigma$ below the MS ($-2\sigma > \Delta\log\rm{SFR} >  -3\sigma$). (7) Fraction of sample that is greater than 3$\sigma$ below the MS ($\Delta\log\rm{SFR} <  -3\sigma$). The COLD GASS fractions represent the average fractions over all 50 representative subsamples. Due to round-off errors, the sums across each row are not exactly equal to 1.
\end{minipage}
\end{table*}

\begin{figure}
	\begin{centering}
		\includegraphics[width=\columnwidth]{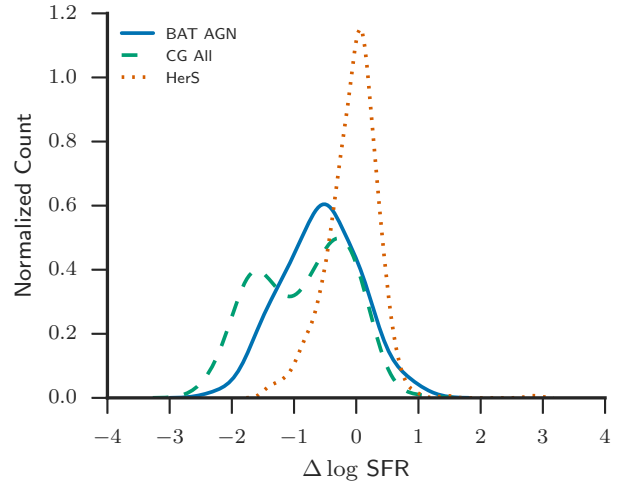}
		\caption{Kernel density estimate (KDE) of the logarithmic distance ($\Delta\log$ SFR) from the main sequence for the BAT AGN, a representative subsample of COLD GASS, and the HerS sample. \label{fig:dlogSFR_all}}
	\end{centering}
\end{figure}
\begin{figure*}
	\begin{centering}
		\includegraphics[width=\textwidth]{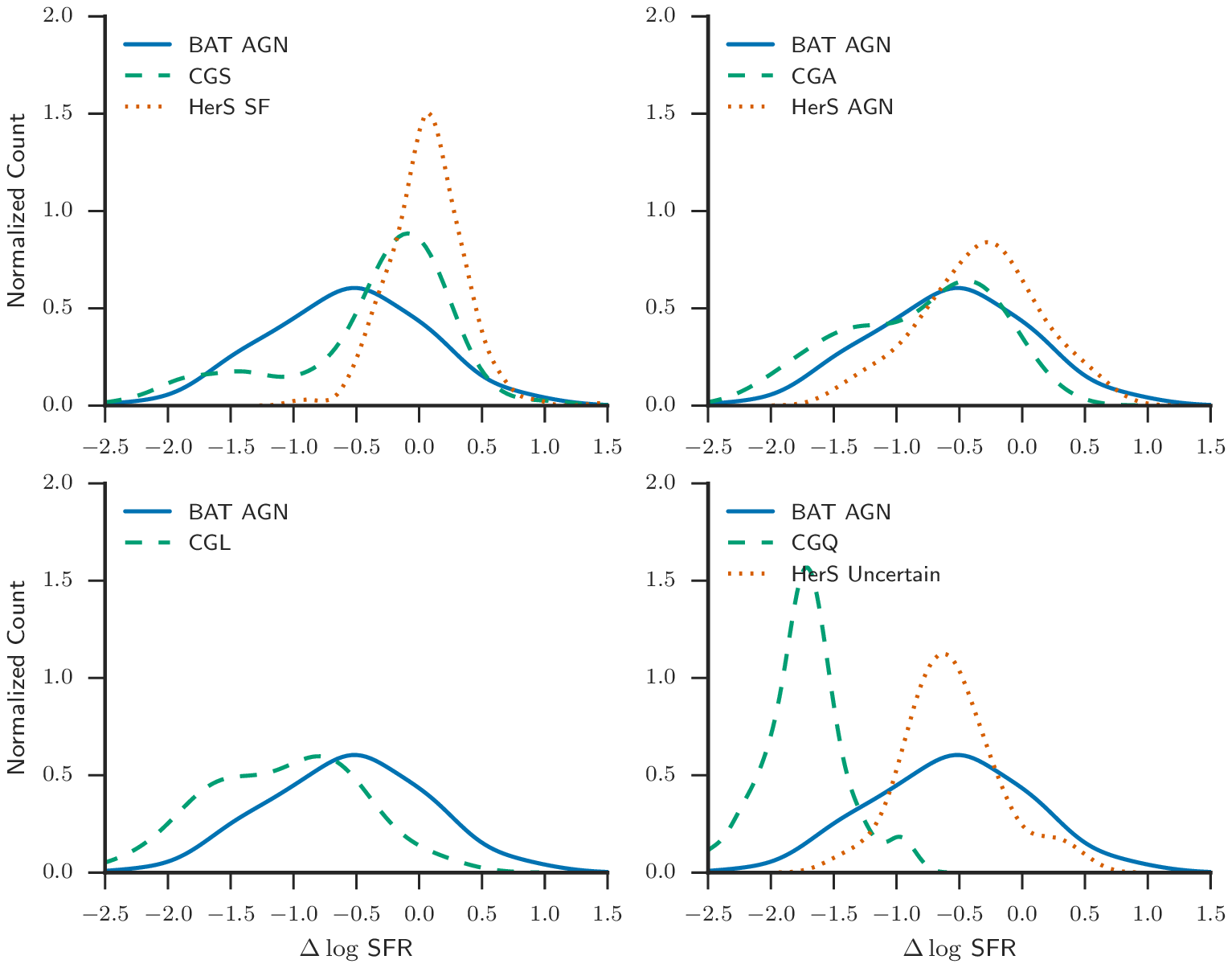}
		\caption{Same as Figure~\ref{fig:dlogSFR_all} but with the COLD GASS subsample split into 4 classifications based on their location in the BPT diagram. CGS: Star-forming galaxies. CGA: AGN and composite galaxies. CGL: LINERs. CGQ: Quiescent galaxies. The HerS sample was split into 3 classifications (Star-forming, AGN and composite, and Uncertain). The BAT AGN galaxies show a similar distribution in $\Delta\log$ SFR as the CGA and HerS AGN sample (upper right). \label{fig:dlogSFR_groups}}
	\end{centering}
\end{figure*}

Figure~\ref{fig:ms_plot} plots the HRS, BAT AGN, a randomly chosen representative subsample of COLD GASS, and the HerS sample on the SFR-\mstar{} plane along with Equation~\ref{eq:ms_hrs_ir} and its scatter. Visually it is clear that a large fraction of the BAT AGN and COLD GASS lie either inside or below the MS. Table~\ref{tab:percentages} quantifies the exact fraction of galaxies in five different regions. The different regions are divided according to $\Delta\log\rm{SFR}=\log\rm{SFR}_{\rm obs} - \log\rm{SFR}_{\rm MS}$ where $\rm{SFR_{obs}}$ is $\rm{SFR}_{\rm IR}$ and $\rm{SFR}_{\rm MS}$ is the SFR expected given the $M_{*}$ of the galaxy using Equation~\ref{eq:ms_hrs_ir}.  $\Delta\log\rm{SFR}$ represents the distance a source is from the main sequence and given the nearly linear MS relation is proportional to specific SFR (sSFR = SFR/\mstar). The five regions are defined as: above the main sequence ($\Delta\log\rm{SFR} > 1\sigma$), inside the main sequence ($1\sigma > \Delta\log\rm{SFR} >  -1\sigma$), 1--2$\sigma$ below the main sequence ($-1\sigma > \Delta\log\rm{SFR} >  -2\sigma$), 2--3$\sigma$ below the main sequence ($-2\sigma > \Delta\log\rm{SFR} >  -3\sigma$), and $3\sigma$ below the main sequence ($\Delta\log\rm{SFR} <  -3\sigma$) with $\sigma$ equal to the observed scatter in the main sequence relationships (i.e. 0.36 dex). We break the ``below'' region in three separate regions to judge how the sample is clustered. If all of the sources below the main sequence are in the $1\sigma$ region, then it could be argued that most of the AGN are main sequence galaxies and simply display a larger scatter. The fractions for the different COLD GASS subsamples are the average fraction over all 50 representative subsamples.

The numbers confirm the visual impression seen in Figure~\ref{fig:ms_plot} that the BAT AGN mainly live inside the main sequence or below it. 28 per cent lie inside the main sequence and 66 per cent lie below it from adding together the three ``below'' regions. Only 5 per cent of the sample is above the main sequence. The sources below the main sequence are also well spread out between the $1\sigma$, $2\sigma$, and $3\sigma$ regions, showing the BAT AGN do not cluster near the edge of the main sequence and in fact a significant percentage (20 per cent) display SFRs more than 3$\sigma$ below what is expected given their stellar masses.

For the COLD GASS sample as a whole, there seems to be a bimodal distribution with 33 per cent of galaxies inside the main sequence and 39 per cent $>3\sigma$ below it. This mirrors what has been extensively seen using optical colors and is another representation of the  split into the blue cloud and red sequence. The BAT AGN do not display the same bimodality and overall have a very different distribution of $\Delta\log\rm{SFR}$ as shown in Figure~\ref{fig:dlogSFR_all}\footnote{We take this opportunity to explain that we choose to represent distributions of values using a Kernel Density Estimate (KDE) rather than a histogram. This is due to the visualization of a histogram being highly dependent on the bin size, number of bins chosen, and the edges of the bins. A KDE represents each point in a data set with a specific kernel and sums all of them together. In this Paper we use a Gaussian kernel. The only tunable parameter is the kernel width for which we use ``Scott's Rule''  \citep{Scott:1992xy}, $width=N^{-1/5}$, where $N$ is the number of data points.} We ran K-S tests to compare the BAT AGN with each COLD GASS representative subsample (see Section~\ref{sec:cold_gass} for a description of the subsamples) and found that 0/50 tests returned a probability $>5$ per cent verifying that the two samples are not drawn from the same parent distribution.

The HerS sample differs from the COLD GASS and BAT AGN samples and is sharply peaked around $\Delta\log\rm{SFR}=0$, signifying most of the galaxies are on the MS. Indeed, 67\% of the total HerS sample lie inside the main sequence, while only 12 per cent are above and 21 per cent below. 
HerS likely does not reach as far below the main sequence as COLD GASS given the parent sample was selected based on a SPIRE 250 \micron{} detection. Using the \citet{Viero:2014jk} 5$\sigma$ depth for SPIRE 250 \micron{} of 65 mJy and assuming a greybody with $T_{dust}=25$ K and $\beta=2.0$, the minimum SFR detectable is $~0.5$ M$_{\odot}$ yr$^{-1}$, much higher than the level reached by COLD GASS and the BAT AGN. Even so, we again ran a K-S test between the BAT AGN and HerS and found a $P_{K-S} << 0.01$ indicating they are drawn from different populations.

It is only when we separate the COLD GASS and HerS sample into their different classifications do we find a similarity. Figure~\ref{fig:dlogSFR_groups} shows the KDEs of $\Delta\log\rm{SFR}$ for the BAT AGN and each COLD GASS and HerS classification. The COLD GASS KDEs were calculated from a randomly chosen representative subsample. The BAT AGN are most similar to the CGA (e.g. the AGN population of COLD GASS), displaying lower values of $\Delta\log\rm{SFR}$ than CGS/HerS SF and higher values than CGL and especially CGQ. In fact the percentages in each region for the BAT AGN and CGA are nearly identical except in the $>3\sigma$ region where there is a larger fraction of CGA. We again ran K-S tests for each of the 50 representative subsamples for COLD GASS and found 48/50 CGA subsamples returned a probability $>5$ per cent indicating that the BAT AGN and CGA are consistent with the same parent population. 0/50 of the CGS, CGL, and CGQ subsamples returned a probability $>5$ per cent of being consistent with the BAT AGN. The K-S test between the BAT AGN and HerS AGN returned a $P_{K-S} =0.002$ while the ones with HerS SF and HerS Uncertain returned $P_{K-S} << 0.001$ and $P_{K-S} = 0.07$. Using the standard 5 per cent cutoff to determine if the samples originate from the same population would indicate that the BAT AGN are most similar to the HerS Uncertain galaxy population. However as we discuss above, this is most likely due to the lower SFR depth reached in HerS. Also the HerS Uncertain classification is a more conservative classification and includes many galaxies that would have been classified using the \citet{Brinchmann:2004bf} system including AGN and LINERs. The HerS AGN still show lower SFRs than the HerS SF sample with 48 per cent inside the MS compared to 78 per cent and 44 percent below it compared to 8 per cent. Therefore, both an optically selected sample of AGN and an ultra-hard X-ray selected sample of AGN display the same property: \textbf{they lie in between a strongly star-forming and quiescent group and suggests that AGN host galaxies are in transition between the two populations.}

We note that while the K-S tests confirm that the BAT AGN and CGA are from the same population, both Table~\ref{tab:percentages} and Figure~\ref{fig:dlogSFR_groups} show that the BAT AGN contain slightly more galaxies with higher SFRs while CGA contains slightly more quiescent galaxies. We hypothesize this is due to the selection method for the two groups. The BAT AGN are X-ray selected, a method that is completely independent from the star-forming properties of the host galaxy while the CGA are selected by optical emission line ratios whose origin can be a mixture of AGN and star-formation. If a galaxy is highly star-forming, optical emission line ratios are more likely to classify it as a star-forming galaxy rather than AGN or even composite (see \citet{Trump:2015db} for biases associated with line ratio selection of AGN).  Indeed, many of the BAT AGN above the main sequence are either involved in a merger or are known starburst galaxies (e.g. Mrk 18, NGC 3079, NGC 7679). 

Because the BAT survey is flux-limited, it is biased against weak AGN, while COLD GASS, due to its selection from SDSS, can reach to lower AGN luminosities. If low-luminosity AGN are more associated with quiescent, early-type galaxies \citep[e.g.][]{Kauffmann:2003mw}, this would explain the larger fraction of the CGA group in the $>3\sigma$ region compared to the BAT AGN. In Section~\ref{subsec:morphologies} we show that the CGA contain a larger fraction of elliptical galaxies than the BAT sample which have lower values of sSFR. 

We also note that the BAT AGN and CGS have essentially the same fraction of galaxies in the $>3\sigma$ below region (16 per cent vs. 14 per cent). This is seen in Figure~\ref{fig:dlogSFR_groups} as the long tail towards low $\Delta\log\rm{SFR}$ for CGS. However, this does not change the general result that the BAT AGN  in general show lower levels of sSFR than the star-forming sample due to the much higher percentages in the $1-2\sigma$ and $2-3\sigma$ below regions. BAT AGN occur in the $1-2\sigma$ region at a $>3$ times higher rate than the CGS and more than 2 times in the $2-3\sigma$ below region. Further 64\% of the CGS occur inside the main sequence compared to only 40\% of the BAT AGN. So over the entire population, AGN are more likely to be found in host galaxies that have lower SFRs than the main sequence.

\subsection{Differences Between Seyfert 1s and 2s}
According to the unified model \citep{Antonucci:1993os,Urry:1995il}, orientation distorts our view of AGN and causes the differences seen between Sy 1s and 2s. Current models invoke an anisotropic dusty and possibly clumpy torus \citep[e.g.][]{Nenkova:2008fr, Honig:2010lr} that absorbs and scatters the nuclear optical/UV/X-ray emission. Two regions of ionized gas produce optical emission lines, one at relatively close distances to the central AGN that produces broad emission lines (i.e. broad line region, BLR) and one at further distances that produces narrow emission lines (i.e. narrow line region, NLR). Under the unified model, Sy 1 galaxies, which display very bright nuclear point sources as wells as broad optical emission lines are viewed along lines of sight through the opening angle of the torus allowing access to the BLR and accretion disk. Sy 2 galaxies on the other hand display weaker or even absent central point sources and only narrow emission lines are viewed through the dusty torus that only allows access to the NLR and obscures emission from the BLR and accretion disk. However, assuming orientation is the only difference means Sy 1s and Sy 2s should display virtually the same host galaxy properties since the torus only affects the very central regions and not the galaxy-wide properties. With our focus on the main sequence, this means Sy 1s and Sy 2s should not separate out in Figure~\ref{fig:ms_plot}.

Indeed, both Figure~\ref{fig:ms_plot} and Table~\ref{tab:percentages} suggest this is the case. Sy 1s compared to Sy 2s in each region show a percent difference of +1, +10, -9, -3, and -0 per cent. Assuming Poisson statistics, these differences are all well within the 1$\sigma$ error bars. Only the 10 and 9 per cent differences for the inside the MS and 1--2$\sigma$ regions have a marginal $1\sigma$ significance. 

\begin{figure}
	\begin{centering}
	\includegraphics[width=\columnwidth]{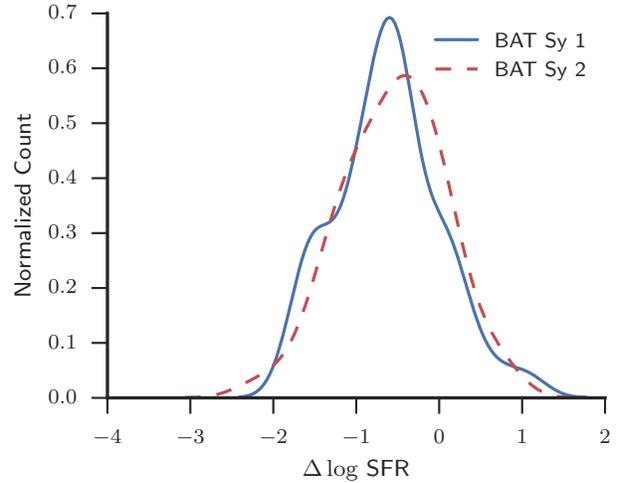}
	\caption{KDE of $\Delta\log$ SFR for Sy 1s (solid, blue line) and Sy 2s (dashed red line) showing the similarity between the two.\label{fig:hist_seyferts}}
	\end{centering}
\end{figure}

To investigate further, we compared the $\Delta\log\rm{SFR}$ distribution for both Sy 1s and Sy 2s and Figure~\ref{fig:hist_seyferts} displays their histograms. Apart from the increased absolute numbers of Sy 2s, Sy 1s and Sy 2s have similar distributions of $\Delta\log\rm{SFR}$. Using a K-S test to test whether they are drawn from the same parent population, we find a p-value of 0.36, again indicating Sy 1s and Sy 2s are similar in terms of their $\Delta\log\rm{SFR}$. This is in agreement with \citet{Koss:2010nr} who found no difference in $u-r$ colors between broad and narrow line AGN. This is in disagreement with previous studies \citep[e.g.][]{Heckman:1989qv, Maiolino:1995nr, Buchanan:2006cq} that found Sy 2s reside in more highly star-forming galaxies. However, these samples and conclusions are based on smaller samples as well as different selection criteria. Samples selected using optical or mid-infrared emission will inherently be influenced by the level of star-formation in the host galaxy and bias samples towards higher star-forming objects, especially for Sy 2's that are fainter at these wavelengths due to obscuration.

\subsection{Correlation of $\Delta\log\rm{SFR}$ with host galaxy and AGN properties}
Given the large percentage of AGN host galaxies below the main sequence compared with normal, main sequence galaxies, we examined the relationship between $\Delta\log\rm{SFR}$ and various AGN and host galaxy properties. Because our entire sample consists of AGN, the immediate reaction is to assume the AGN has influenced star formation in the host galaxy through some mechanism and slowed it down. From this scenario, the expectation is for more powerful AGN to have a greater effect on the host galaxy and occur further from the main sequence. To test this, we binned the sample according to the regions described in Section~\ref{subsec:agn_ms_loc}. Within each bin we calculated the mean 14--195 keV luminosity (\lx) since \cite{Winter:2012yq} found the 14--195 keV luminosity to be a very good probe of the overall bolometric luminosity and strength of the AGN.

\begin{figure*}
	\begin{center}
	\includegraphics[width = \textwidth]{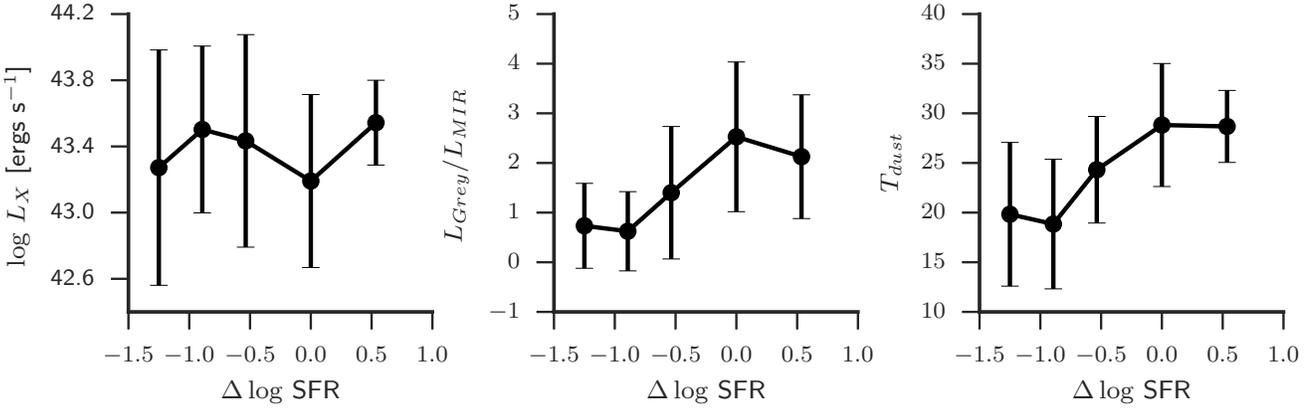}
	\caption{Correlations between different properties of the galaxies/AGN as a function of $\Delta\log\rm{SFR}$. \textit{left}: 14--195 keV luminosity, \textit{middle}: ratio of the MIR power law luminosity and greybody luminosity, \textit{right}: Dust temperature. For each property, we binned the sources according to whether they were above, inside, or below the main sequence (split into 3 separate regions). For each MS region we calculated the mean AGN/galaxy property. Error bars are the standard deviation within each bin.\label{fig:sSFR_prop}} 
	\end{center}
\end{figure*}

The top left plot of Figure~\ref{fig:sSFR_prop} shows the resulting relationship between $\Delta\log\rm{SFR}$ and \lx.  We found \textbf{no} clear correlation between the strength of the AGN and $\Delta\log\rm{SFR}$. Over the entire range of $\Delta\log\rm{SFR}$, the mean \lx{} only changes by $<0.4$ dex with a large spread in each region. We calculated the Spearman rank correlation coefficient ($\rho_{s}$) and used bootstrap analysis to determine the 95 per cent confidence interval. We found $\rho_{s} = -0.1$ with a 95 percent confidence interval of  -0.3--0.1 consistent with $\rho_s = 0$. This would seem to argue against the AGN having any effect on star-formation in the host galaxy. However \citet{Hickox:2014yq} argues that AGN variability will smear out any intrinsic correlations between star-formation and AGN activity due to the much shorter timescales of AGN activity compared to star-formation. The ultra-hard X-rays used to calculate the luminosity for the BAT AGN presumedly originate very near the SMBH and represent an instantaneous strength while the SFRs are averaged over $\sim$100 Myr. The null correlation and large scatter we see between \lx{} and $\Delta\log\rm{SFR}$ then is most likely a product of the large variability that AGN typically exhibit.

The middle plot indicates there is a positive correlation between $L_{\rm Grey}/L_{\rm MIR}$ and $\Delta\log\rm{SFR}$. $L_{\rm Grey}/L_{\rm MIR}$ is the ratio of the luminosity of the greybody component to the luminosity of the MIR power-law component used in our SED model. The MIR power law luminosity strongly correlates with the AGN luminosity (Shimizu et al. 2015, in preparation) and is assumed to be produced by the AGN, while the greybody luminosity is assumed to be a product of star formation. The ratio of their luminosities is a measure of which component dominates the SED. Far below the MS, the IR SED should be dominated by the AGN, while the IR SEDs of MS galaxies as well as those above the MS are dominated by star formation. We expect this trend due to the null correlation of \lx{} with $\Delta\log\rm{SFR}$. Since $L_{\rm MIR}$ is strongly correlated with \lx, a decrease in $\Delta\log\rm{SFR}$ is then mostly due to a decrease in SFR which was directly calculated from $L_{\rm grey}$. We again calculated $\rho_{s}$ finding $\rho_{s} =  0.6$ [0.46--0.7] where the range in brackets is the 95 per cent confidence interval determined using a bootstrap analysis. This strongly suggests a real positive correlation between $L_{\rm Grey}/L_{\rm MIR}$ and $\Delta\log\rm{SFR}$ and confirms that far below the main sequence the IR SED is most likely dominated by the AGN.

The greybody dust temperature also suggests a positive correlation with $\Delta\log\rm{SFR}$ with galaxies at larger $\Delta\log\rm{SFR}$ having a higher temperature (Figure~\ref{fig:sSFR_prop}, right). This has been observed before \citep[e.g.][]{Magnelli:2014sf} and can easily be explained given that an increase in SFR increases the number of OB stars that produce the UV photons to heat the dust. We found $\rho_{s} =  0.6$ [0.45--0.7], very similar to the correlation with $L_{\rm Grey}/L_{\rm MIR}$ showing that the true property determining an AGN host galaxy's location within the main sequence diagram is star-formation rather than the strength of the AGN.

\begin{figure}
	\includegraphics[width=\columnwidth]{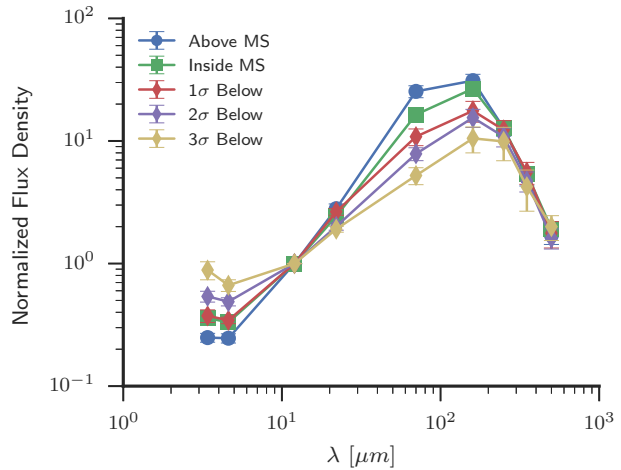}
	\caption{Mean SED of the sources in each region of the main sequence plot. The individual SEDs were first normalized to the 12 \micron{} flux density. Error bars represent the standard error of the mean normalized flux.\label{fig:mean_SED}}
 \end{figure}

Both of these effects can also be seen in Figure~\ref{fig:mean_SED} where we plot the mean SEDs for each region after normalizing to the 12 \micron{} flux density. Because some of the observed SEDs contain upper limits especially at the longest wavelengths, we use the Kaplan-Meier product-limit estimator \citep{Feigelson:1985lr}, a maximum likelihood estimate of the distribution function, to calculate the mean and its standard error. Using only the detected flux densities would bias the mean towards larger values, with an increasing bias at longer wavelengths as the number of upper limits increases. This would then lead to an artificial flattening of the SED. 

The peak of the SED is seen to shift to longer wavelengths as sSFR decreases, indicating colder temperatures (i.e. Wien's Displacement Law) while the overall amplitude decreases along with a flattening of the slope of the SED between 4.6 and 160 \micron. In their analysis of \textit{Spitzer} IRS spectra for Seyfert galaxies, both \citet{Wu:2009pt} and \citep{Baum:2010kh} found that AGN host galaxies with lower amounts of star formation display bluer SEDs in 15--40 \micron{} regime, in agreement with our mean SEDs and an increase in the AGN contribution. We also see a general increase in the W1/W2 \textit{WISE} color moving towards lower sSFR, an indication that the host galaxy (in particular older stars) begins to contribute to the MIR, similar to what is observed in many quenched galaxies. The long wavelength IR regime, though, seems to be completely unaffected by the sSFR. All five of the SEDs essentially display the same slope and relative flux density at 250, 350, and 500 \micron. Indeed a comparison of the BAT AGN SPIRE colors with the HRS colors shows there is little difference between the two samples (Shimizu et al 2015, in preparation), verifying the same process (i.e. star formation) is producing most of the long wavelength emission. 

\subsection{Host Galaxy Morphology}\label{subsec:morphologies}

\citet{Koss:2011vn} closely analyzed the host galaxy morphologies of the BAT AGN, finding that at all stellar masses a larger percentage of AGN are hosted by spiral galaxies compared to a matched sample of normal galaxies. Using these morphologies, we can assess if we observe a change in the host galaxy morphology as a function of sSFR. \citet{Koss:2011vn} classified the BAT AGN into three categories: spirals, ellipticals, and intermediate based on the results from the Galaxy Zoo Project \citep{Lintott:2008qv}. Each galaxy was independently classified numerous times by the public. A spiral or elliptical morphology was chosen for the galaxy if $>80$ per cent people selected the type, or else intermediate was chosen. Mergers were defined in the same way as \citet{Patton:2008ty} and \citet{Koss:2010nr}, requiring a projected distance of at most 30 kpc and a radial velocity difference of $<500$ km s$^{-1}$ between the galaxy and its companion. We used the same method for the COLD GASS sample as well for the spiral/elliptical/intermediate classification. Merger classifications for the COLD GASS sample were determined in \citet{Saintonge:2012uq} where they visually classified each galaxy as a merger/interaction if it had a nearby companion ($<1'$) and/or evidence of a disturbed morphology, tidal tails, etc (see Appendix A of \citet{Saintonge:2012uq}). Each galaxy that looked like a merger was given a merger rating from 2--5 with 2 representing galaxies that only had a nearby companion and a 5 representing galaxies with very strong signs of a merger. After looking through the images of the COLD GASS mergers we decided to exclude all of the galaxies with a merger rating of 2 because they would not have been classified as a merger using the method from \citet{Koss:2011vn}. We chose to only compare the BAT AGN morphologies with the COLD GASS sample given COLD GASS's better mass completeness and larger SFR depth.

\begin{figure*}
	\begin{centering}
		\includegraphics[width=\textwidth]{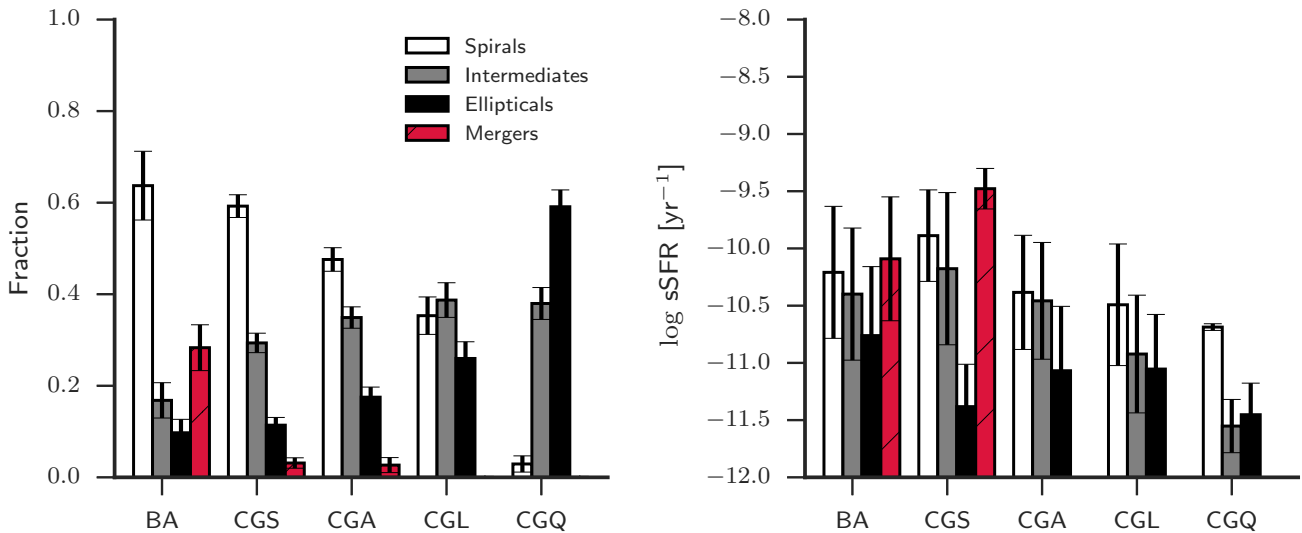}
		\caption{\textit{Left:} The distribution of spirals, intermediates, and ellipticals for all of the galaxy samples. For the COLD GASS samples, the ractions represent the mean fraction of all 50 representative subsamples. $1\sigma$ error bars for the BAT AGN (BA) were calculated assuming Poisson statistics while the error bars for the COLD GASS samples represent the standard deviation of the fractions for the 50 representative subsamples. \textit{Right:} The average sSFR for the different morphologies within each sample. For the COLD GASS samples we randomly chose one representative subsample which for the CGA in this case did not contain any mergers. Error bars are the standard deviation of sSFR within each group. \label{fig:morph_breakdown}}
	\end{centering}
\end{figure*}

In the ``Total'' column of Table~\ref{tab:morph_ms} and Figure~\ref{fig:morph_breakdown} we outline the total fraction of galaxies that are spirals, intermediates, and ellipticals over the whole sample. Spiral galaxies dominate ($\sim60$ per cent) the population of both the BAT AGN and CGS similar to what was found in \citet{Koss:2011vn} whereas for the CGA they represent $\sim50$ per cent of the population and for the CGL they are 35 per cent. There are virtually no spiral galaxies in the quiescent group (CGQ). 

\begin{figure*}
	\begin{centering}
		\includegraphics[width=\textwidth]{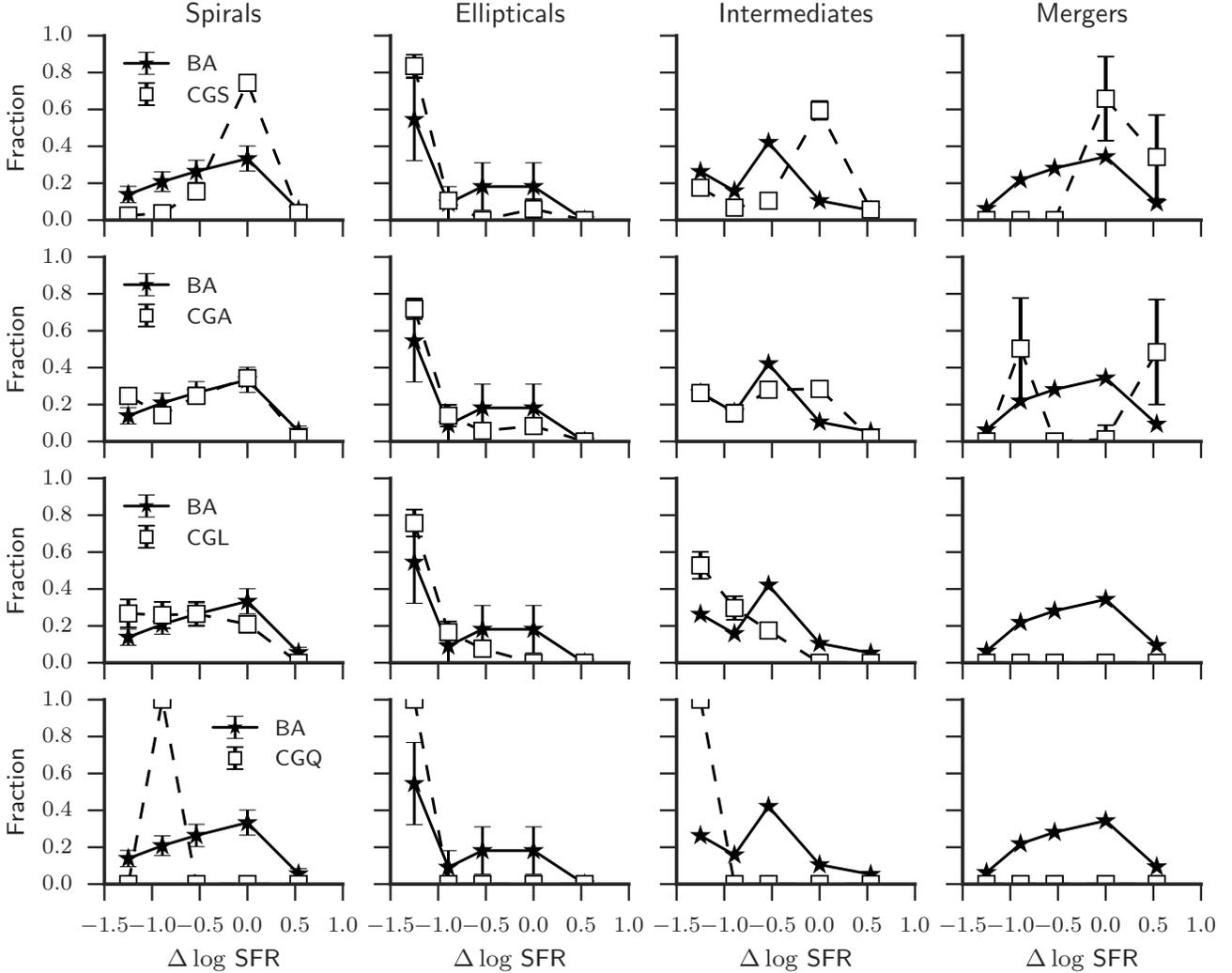}
		\caption{Comparison of the fraction of spirals, ellipticals, intermediates, and mergers that are in each main sequence region for the BAT AGN (BA) and COLD GASS samples (CGS, CGA, CGL, and CGQ). The fractions for the COLD GASS samples are the average fractions over the 50 representative subsamples. Errors for the BAT AGN are calculated assuming Poisson statistics while the errors for the COLD GASS samples are the standard deviation of the fraction for the 50 representative subsamples.\label{fig:morph_breakdown_group} }
	\end{centering}
\end{figure*}

On the other hand the CGQ are dominated by elliptical galaxies ($\sim60$ per cent). The CGL are $\sim26$ per cent elliptical, CGA 20 per cent, and the BAT AGN and CGS 10 per cent elliptical. Overall in terms of the whole population the BAT AGN more closely resemble the morphology distribution of the CGS rather than the CGA. However a closer look at the sSFR for each morphology shows that the BAT AGN and CGA are more similar. This is shown in Figure~\ref{fig:morph_breakdown}, right panel, where we plot the average sSFR for each morphology. BAT AGN spirals and intermediates show decreased levels of sSFR compared to the CGS and more in line with the CGA. If we examine the distribution of spirals, ellipticals, and intermediates as shown in Figure~\ref{fig:morph_breakdown_group}, we can again see that the BAT AGN and CGA are almost identical in terms of the fraction of spirals and ellipticals in each MS region. The biggest difference seems to occur with intermediates; however there is much uncertainty concerning the nature of intermediates given they are simply defined as objects where there was no consensus on whether it was a spiral or elliptical.

Interestingly for both the BAT AGN and CGA, ellipticals show increased levels of sSFR compared to those within the CGS and CGQ groups. This could be an indication that AGN within early-type galaxies actually stimulate star-formation rather than quench it. With such large error bars, though, the significance of this result is unclear.

In agreement with \citet{Koss:2010nr} we find that the merger fraction of the BAT AGN (28 per cent) is much higher than that of the CGS (3 per cent) and CGA (3 per cent). There are no mergers in the CGL and CGQ samples. Mergers have long been known to increase the SFR in the interacting galaxies \citep{Sanders:1988fk} as evidenced by the majority of ultra-luminous infrared galaxies (ULIRGS) involved in one. \citet{Elbaz:2011qe} found that the nearly all of the local galaxies above the main sequence are IR-compact starbursts that were most likely triggered by a merger. Thus, we expect to see a large fraction of mergers occurring above the main sequence. Indeed, this is the case  especially for the CGS where 34 per cent of the mergers are above the MS and the rest are inside the MS. Further for both the BAT AGN and CGS, mergers have the greatest average sSFR (Figure~\ref{fig:morph_breakdown}). This could also explain the extremely low merger fraction for the CGA. If mergers are associated with highly star-forming galaxies, then emission line ratios would indicate a star-forming galaxy rather than an AGN as any AGN signatures would be overwhelmed. With ultra-hard X-rays though, we can peer through the obscuring gas and detect the AGN.

\begin{figure}
\includegraphics[width=\columnwidth]{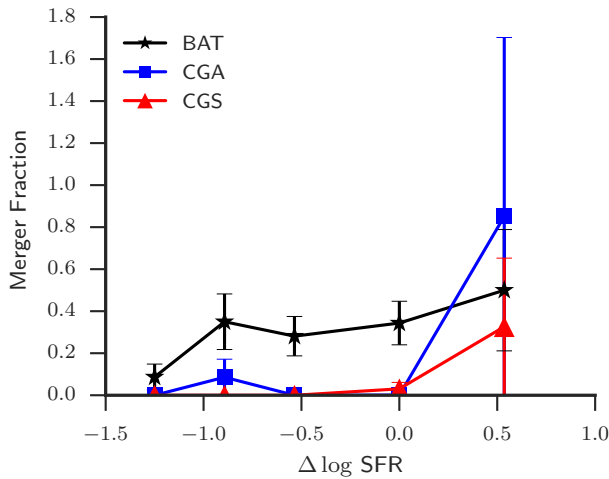}
\caption{ Fraction of galaxies in each MS region that are mergers for the different samples.. Errors for the BAT AGN (black stars) are calculated assuming Poisson statistics for the merger fraction. Errors for the CGA (blue squares) and CGS (red triangles) are the standard deviation of the fraction over all 50 representative subsamples. \label{fig:merge_frac}}
\end{figure}

In Figure~\ref{fig:merge_frac}, we plot the fraction of all galaxies of a sample in a MS region that is a merger. This is different than Figure~\ref{fig:morph_breakdown_group} where we plot the fraction of all mergers that is in each region. Here we can see that the merger incidence rate rises with $\Delta\log\rm{SFR}$ as expected. Above the MS, the CGA show the highest merger rate at roughly 80 per cent, but this is based on small number statistics and is reflected in the large error bars. The BAT AGN have a 50 per cent merger rate and CGS are at 30 per cent. For all three samples the above the MS region represents the largest merger rate. Inside, and below the MS the BAT AGN clearly show much larger merger rates than that of CGS and CGA. In fact the merger fraction for the BAT AGN is roughly flat from above the MS to the 2--3$\sigma$ region. This could be evidence that mergers and interactions are important in triggering AGN at low redshift even before the star burst ignites and those below the MS are in fact moving up the SFR-\mstar{} plane. Another explanation is that these are late stage mergers and have moved past the initial burst of star formation and are slowly falling off the main sequence. A confirmation of this would be if all of the mergers in the $>1\sigma$ region are in the early stages whereas the ones above the MS are near coalescence. In either case, the BAT AGN mergers occur across nearly the whole SFR-\mstar{} plane at a high rate compared to local non-active galaxies and optically selected AGN. This is in disagreement with recent higher-redshift studies of the merger rate in X-ray selected galaxies \cite[e.g.][]{Kocevski:2012lr, Villforth:2014qy} that find no difference in the merger rates between AGN and non-AGN galaxies.

\begin{table*}
\begin{centering}
\begin{minipage}{110mm}
	\caption{Host Galaxy Morphology Distribution in SFR-\mstar{} Plane}
	\label{tab:morph_ms}
	\begin{tabular}{@{}lcccccc}
		\hline
		Sample & Total & Above MS & Inside MS & $1-2\sigma$ Below & $2-3\sigma$ Below & $> 3\sigma$ Below \\
		(1) & (2) & (3) & (4) & (5) & (6) & (7) \\
		\hline
		\textbf{BAT AGN} \\
		Spirals & 0.64 & 0.06 & 0.33 & 0.26 & 0.21 & 0.14 \\
		Ellipticals & 0.10 & 0.00 & 0.18 & 0.18 & 0.09 & 0.55 \\
		Intermediates & 0.17 & 0.05 & 0.11 & 0.42 & 0.16 & 0.26 \\
		Mergers & 0.28 & 0.09 & 0.34 & 0.28 & 0.22 & 0.06 \\
		\\
		\textbf{CG All} \\
		Spirals & 0.43 & 0.03 & 0.52 & 0.20 & 0.12 & 0.13 \\
		Ellipticals & 0.23 & 0.00 & 0.03 & 0.03 & 0.08 & 0.86 \\
		Intermediates & 0.34 & 0.03 & 0.28 & 0.16 & 0.13 & 0.41 \\
		Mergers & 0.02 & 0.41 & 0.40 & 0.00 & 0.19 & 0.00 \\
		\\
		\textbf{CGS} \\
		Spirals & 0.59 & 0.04 & 0.74 & 0.16 & 0.04 & 0.03 \\
		Ellipticals & 0.11 & 0.00 & 0.06 & 0.00 & 0.11 & 0.84 \\
		Intermediates & 0.29 & 0.06 & 0.60 & 0.11 & 0.07 & 0.18 \\
		Mergers & 0.03 & 0.34 & 0.66 & 0.00 & 0.00 & 0.00 \\
		\\
		\textbf{CGA} \\
		Spirals & 0.48 & 0.02 & 0.34 & 0.25 & 0.14 & 0.25 \\
		Ellipticals & 0.18 & 0.00 & 0.08 & 0.06 & 0.14 & 0.72 \\
		Intermediates & 0.35 & 0.02 & 0.29 & 0.28 & 0.15 & 0.26 \\
		Mergers & 0.03 & 0.49 & 0.01 & 0.00 & 0.50 & 0.00 \\
		\\
		\textbf{CGL} \\
		Spirals & 0.35 & 0.00 & 0.21 & 0.27 & 0.26 & 0.27 \\
		Ellipticals & 0.26 & 0.00 & 0.00 & 0.08 & 0.17 & 0.76 \\
		Intermediates & 0.39 & 0.00 & 0.00 & 0.18 & 0.30 & 0.53 \\
		Mergers & 0.00 & 0.00 & 0.00 & 0.00 & 0.00 & 0.00\\
		\\
		\textbf{CGQ} \\
		Spirals & 0.03 & 0.00 & 0.00 & 0.00 & 1.00 & 0.00 \\
		Ellipticals & 0.59 & 0.00 & 0.00 & 0.00 & 0.00 & 1.00 \\
		Intermediates & 0.38 & 0.00 & 0.00 & 0.00 & 0.00 & 1.00\\
		Mergers & 0.00 & 0.00 & 0.00 & 0.00 & 0.00 & 0.00\\
		\hline
	\end{tabular}
\end{minipage}
\end{centering}
\end{table*}

\section{Discussion}
\subsection{Selection Effects and Model Dependence\label{subsec:sel_effects_mod_depend}}
Due to the flux-limited nature of our parent AGN sample, there is a strong dependence on the X-ray luminosity (i.e. AGN luminosity) with distance, limiting the inclusion of low luminosity AGN especially at higher redshifts. However this would only affect our results if they were related to either the SFR or \mstar{} of the host galaxy. X-ray luminosity has been shown to only have a weak correlation with SFR \citep[e.g][]{Silverman:2009lq, Shao:2010fp, Mullaney:2012gf, Rosario:2012fr}. Furthermore, the correlations are positive implying the absence of low AGN luminosity in our sample only removes objects with even lower SFR that would enhance our result that a large fraction of AGN lie below the main sequence. \citet{Mullaney:2012gf} found a very weak correlation of intrinsic AGN luminosity with stellar mass where a two order increase in AGN luminosity results in only a factor 2 greater stellar mass. This combined with the weak correlation of SFR shows that we are not biased towards low sSFR galaxies in our sample.

The choice of model to fit the SEDs of our galaxies could greatly influence the results, given that the SFRs are directly calculated from the measured FIR luminosity. It is possible that we are assigning too much of the 8--1000 \micron{} luminosity to the AGN and underestimating the SFR. We have tested this effect in two ways. First, we implemented the SED fitting routine, DECOMPIR, described and developed by \citet{Mullaney:2011yq}. DECOMPIR utilizes a set of 5 host galaxy templates and an intrinsic AGN template to fit the IR SED's of galaxies. Although there is the option to let the parameters of the AGN template vary, given our sample, we only used the mean intrinsic AGN SED found in \citet{Mullaney:2011yq}. Each BAT AGN was fit with each host galaxy template allowing the normalizations of both the AGN and host galaxy to vary. The host galaxy template that resulted in the minimum $\chi^{2}$ was chosen as the best fit. SFRs were then calculated by integrating the best fit host galaxy SED from 8--1000 \micron{} and using Equation~\ref{ir_sfr}. Using the DECOMPIR SFRs combined with the same stellar masses and main sequence relation (defined by Equation~\ref{eq:ms_hrs_ir}) we find 6, 35, 28, 14, and 18 per cent of the AGN in the Above MS, Inside MS, $1-2\sigma$ Below, $2-3\sigma$ Below, and $> 3\sigma$ Below regions respectively. These fractions are extremely similar to those using the \cite{Casey:2012jl} model (Table~\ref{tab:percentages}) with only a slightly higher incidence of AGN inside the main sequence and lower number below it. 

The second way we tested the model dependency was by using just the monochromatic 160 \micron{} luminosity as a pure SFR indicator. The 60 or 70 \micron{} luminosity has been used extensively in the literature as a SFR indicator; however, based on the modeling of the BAT sample, 70 \micron{} is a poor choice due to the AGN contribution especially at higher AGN luminosities \citep{Melendez:2014yu}. 160 \micron{} however seems to be relatively free from any AGN contribution. Therefore we simply used the 160 \micron{} luminosity to convert to a SFR using the relation from \citep{Calzetti:2010cj} for both the HRS sample and BAT AGN and recalculated the main sequence based only on the HRS since we do not have 160 \micron{} data for the HerS sample. This results in 6, 37, 25, 12, and 20 per cent of the BAT AGN in the Above MS, Inside MS, $1-2\sigma$ Below, $2-3\sigma$ Below, and $> 3\sigma$ Below regions respectively, comparable to both the values in Table~\ref{tab:percentages} and using DECOMPIR. This confirms our main result that a large fraction of AGN lie below the main sequence and is not a product of over-subtracting the AGN component of the SED. Three separate techniques for estimating the SFR agree that $>50$ per cent of AGN have lower sSFRs than normal galaxies for the same stellar mass.

\subsection{Comparison with previous studies}
We have substantiated that our results are not due to selection effects or are model dependent. The BAT AGN, COLD GASS AGN, and HerS AGN span a large range in sSFR that extends over two orders of magnitude. Many previous studies have suffered from low detection rates ($<40$ per cent) in the FIR and needed to resort to stacking techniques \citep[e.g.][]{Mullaney:2012gf} to achieve a reasonable dynamic range. Our \herschel-BAT AGN sample provides a unique opportunity to compare to previous work and whether assumptions made remain valid. In particular we will focus on the work from \citet{Mullaney:2012gf} who used deep \herschel{} observations of the GOODS-South and GOODS-North fields to study the star-forming properties of AGN selected from 2 Ms and 4Ms \textit{Chandra} Deep Field South and North. Their AGN sample spans the same luminosity range as the BAT AGN, but a much higher redshift range ($0.5<z<3$) with an overall detection fraction in the FIR of $\sim40$ per cent. To enhance their sample to low redshifts, they utilize a sample of the BAT AGN, however with much poorer FIR photometry from \textit{IRAS} and again only a detection fraction of $40$ per cent compared to our 95 per cent and 83 per cent detection fraction at 70 and 160 \micron{} respectively, the two wavelength bands closest to the longest \textit{IRAS} bands. 

For the FIR undetected AGN in the GOODS fields, they utilized stacking analysis to find the weighted average sSFR in three different redshift bins. Using a Monte Carlo approach and assuming a lognormal distribution of sSFR, they estimated 15 per cent of all (FIR detected and undetected) AGN reside in quenched galaxies, 79 per cent are in main-sequence galaxies, and 7 per cent in starburst galaxies. While the starburst percentage (i.e. above the main sequence) is comparable to the fraction presented here, the main sequence and quenched (i.e. below the main sequence) percentages seem to be in disagreement. However this is merely due to a difference in definition of being inside or below the main sequence. \citet{Mullaney:2012gf} define starbursts as galaxies that have sSFRs $>3$ times that of main sequence galaxies and quenched galaxies as ones that have 10 per cent the sSFR of main sequence galaxies. All others are then considered normal star-forming galaxies which include transitioning galaxies in the ``green valley.'' Using these criteria for our BAT sample, we then find 14 per cent starburst galaxies, 73 per cent main sequence galaxies, and 13 per cent quenched galaxies. While we seem to see a factor of 2 greater incidence of starburst galaxies, the fraction of main sequence galaxies and quenched galaxies are completely in line with \citet{Mullaney:2012gf}.

To further check the agreement between the two studies, we simulated samples of AGN at higher redshifts by assuming our BAT AGN sample is representative of the AGN population at all redshifts. To simplify the process, we also assumed every AGN has a host galaxy SED exactly that of the SB5 template from DECOMPIR at $z=0$. The template was then scaled to have the same observed 160 \micron{} flux. For each redshift bin probed by \citet{Mullaney:2012gf} (0.5--1.0, 1--2, and 2--3) we simulated 1000 samples of AGN by randomly assigning a redshift from that bin to each of the AGN in our BAT sample. After a redshift was assigned, we simulated the increase in SFR with increasing redshift using Equation 13 from \citep{Elbaz:2011qe}. From this, we obtained a multiplicative factor to boost the IR SED of the AGN for a specific redshift. Finally, after redshifting the SED, we employed the same detection wavelengths (100 \micron{} for $0.5<z<1.5$ and 160 \micron{} for $1.5<z<3.0$) and thresholds (0.8 mJy for 100 \micron{} and 2.4 mJy for 160 \micron{}) to determine whether the source would be detected.  The detection fractions for all 1000 sets of AGN were averaged to find the estimated detection fraction for each redshift bin.

Overall for the entire redshift range of 0.5--3 we found a detection fraction of $\sim35$ per cent, very near to the observed one of 42 per cent. However when splitting up into the redshift bins we find detection fractions of 71, 30, and 6 per cent compared to 60, 42, and 29 per cent found by \citet{Mullaney:2012gf} in redshift bins of 0.5--1.0, 1--2, and 2--3 respectively. This could be a sign of an evolution of AGN host galaxies where at higher redshifts, more AGN live in main sequence galaxies than what is seen at low redshift. The discrepancy could also be caused by the lack of an AGN contribution used in our simulations. At 100 and 160 \micron{}, there is very little contamination from the AGN, but at the rest frame wavelengths being probed (40--67 \micron) it could have a profound effect. To test this we repeated the simulations, adding in the average AGN SED from DECOMPIR such that 25 per cent of the observed frame 100 or 160 \micron{} is due to the AGN. Doing this increases the detection fraction to 77, 38, and 10 per cent in the same redshift bins, much closer in the middle bin but still far off the 30 per cent in the highest redshift bin. However, we have not included any AGN luminosity evolution with redshift. \citet{Ueda:2014fv} find that \lx{} is a strong function of redshift especially between $z=0-2$. This would increase the fluxes especially in the higher redshift bins and subsequently increase the detection fractions. 

\subsection{Comparison with SDSS sample}
All three AGN samples, one X-ray selected (BAT AGN) and two optically selected (CGA and HerS AGN) have reduced levels of star-formation compared to the main sequence. All of these samples, however, are relatively small in number so we decided to test a much larger sample of both star-forming and AGN galaxies from the SDSS.

The SDSS DR7 spectroscopic catalog mentioned previously contains 818,333 unique galaxies with optical spectra that were analyzed in a consistent manner by the MPA-JHU team \citep{Brinchmann:2004bf}. SFRs and stellar masses were measured for every galaxy. Stellar masses are based on fits to the five SDSS photometry using the technique described in \citet{Salim:2007rm}. SFRs were derived in two separate ways. For star-forming galaxies, the SDSS spectra were fit using the \citet{Charlot:2001rt} models with an additional aperture correction to account for light outside the fiber \citep{Salim:2007rm}. For the other classifications, emission lines are not reliable either due to low S/N or unknown AGN contribution (composite and AGN dominated). The SFRs for these groups were estimated using the 4000 \AA break (\emph{D$_{n}$4000}) which has been shown to correlate with sSFR, albeit with large scatter \citep{Brinchmann:2004bf}. 

\begin{figure}
	\begin{centering}
		\includegraphics[width=\columnwidth]{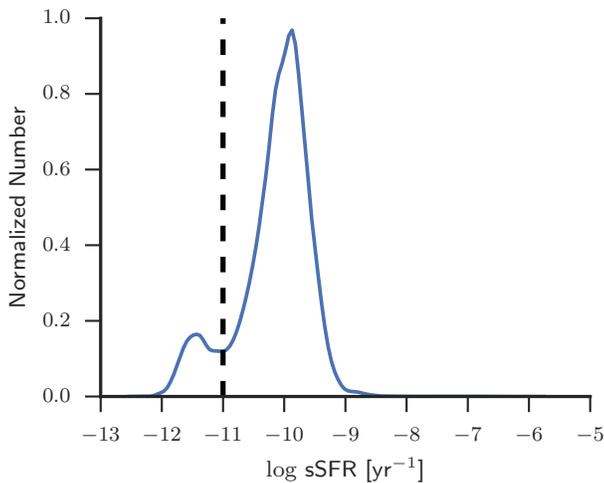}
		\caption{KDE of the sSFR for the SDSS DR7 sample. There is a clear bi-modality in sSFR defining the two populations of actively star forming and passive galaxies with a transition around $10^{-11}$ yr$^{-1}$ (dashed line).\label{sdss_ssfr}}
	\end{centering}
\end{figure}

We restricted the SDSS sample to only those galaxies with well measured redshifts ($z_{conf} > 0.9$) and in the redshift range $0.01<z<0.05$. The upper limit matches the upper redshift of our BAT AGN sample and the lower limit is to avoid very nearby galaxies that the SDSS pipeline can shred into multiple sources. We further included an absolute magnitude cut of $M_{z} < -19.5$ to ensure mass completeness and only used galaxies with well measured SFRs ($\rm{sfr\_flag = 0}$). These restrictions resulted in 21,091 star-forming galaxies; 8,189 low-SFR galaxies; 12,190 composite; and 9,421 AGN-dominated systems. We combined the star-forming and low-SFR galaxies to form the SDSS normal galaxy sample. Figure~\ref{sdss_ssfr} shows the distribution of sSFR for the SDSS normal galaxies. There is a clear bi-modality in sSFR that matches the bi-modality seen with optical colors and in the COLD GASS sample. The threshold between the two populations occurs at $\rm{sSFR}\sim10^{-11} \rm{yr}^{-1}$ shown by the vertical dashed line. We use this threshold to split the SDSS normal galaxy sample into an actively star-forming population and passive population, hereafter referred to as SDSS$_{\rm{SF}}$ and SDSS$_{\rm{passive}}$. 

\begin{figure*}
	\begin{centering}
		\includegraphics[width=\textwidth]{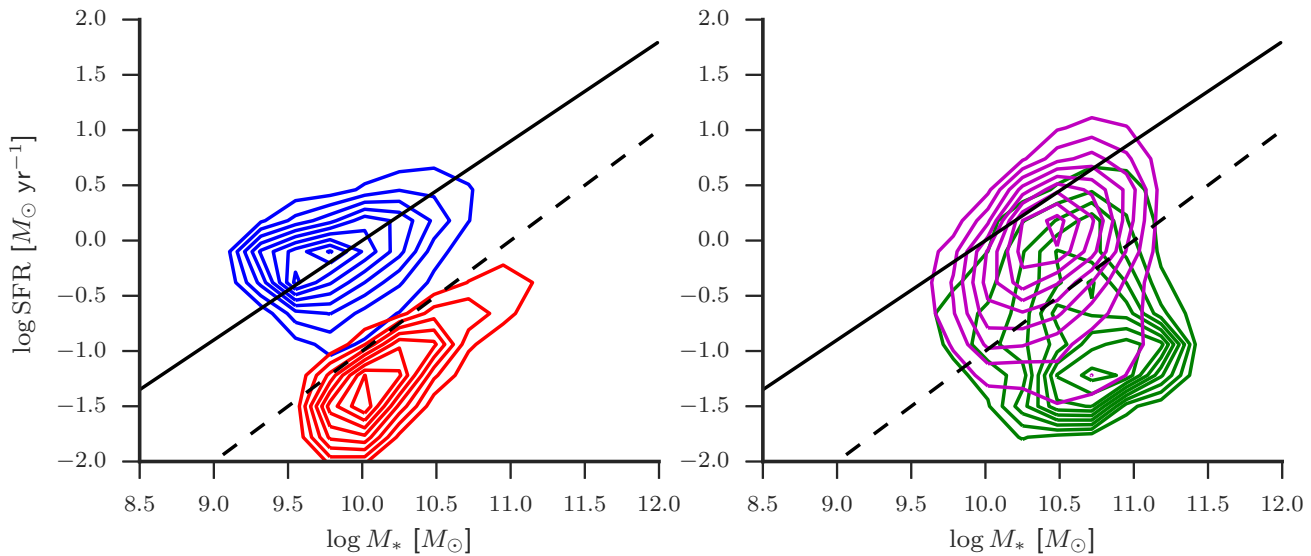}
		\caption{SFR--$M_{*}$ diagram for the spectroscopic SDSS DR7 sample. {\it left panel:} SDSS non-AGN galaxies split into ``SF'' (blue contours) and ``passive'' (red contours) based on the sSFR. The black line indicates the sSFR$=10^{-11}$ yr$^{-1}$ threshold to split the sample into ``active'' and ``passive''. {\it right panel:} Same as left panel with contours representing the SDSS composite (magenta contours) and AGN-dominated galaxies (green contours). Contours enclose 10--90 per cent of the specified sample in increments of 10 per cent. \label{sdss_bat_agn_ms}}
	\end{centering}
\end{figure*}

In Figure~\ref{sdss_bat_agn_ms} we plot the number density of the SDSS galaxies on the SFR-$M_{*}$ diagram separated into the various classifications.  In the left panel of Figure~\ref{sdss_bat_agn_ms}, we plot the SDSS$_{SF}$ galaxies (blue contours) along with the SDSS$_{passive}$ galaxies. The black solid line is the MS relation found by \citet{Peng:2010wd} using the same data while the dashed line corresponds to a constant sSFR$=10^{-11}$ yr$^{-1}$. The SDSS$_{SF}$ galaxies follow the \citet{Peng:2010wd} MS relation just as the HRS galaxies follow their own MS relation with the passive galaxies far below. In the right panel of Figure~\ref{sdss_bat_agn_ms} we plot the SDSS composite (magenta contours) and AGN-dominated galaxies (green contours). The same effect is seen as when comparing the BAT AGN, CGA, and HerS AGN with the main sequence, CGS, and HerS SF. \textbf{AGN host galaxies systematically have lower rates of star formation than normal star-forming galaxies.} SDSS AGN and composites, BAT AGN, CGA, and HerS AGN definitively lie in between the actively star-forming and completely quenched population signaling these galaxies are possibly transitioning from one stage to the next. 
\subsection{Implications for Galaxy Evolution and AGN Feedback}
In this Paper, we have rigorously shown that Seyfert galaxies display lower levels of star formation than that expected from the main sequence by comparing normal galaxy samples with an ultra-hard X-ray selected AGN sample using the same methods for measuring the SFR and stellar mass. Optically selected AGN from the COLD GASS and HerS sample also show the same effect even though their SFRs were measured using a different method. Extending the comparison to large numbers with the SDSS further emphasizes the difference between AGN and non-AGN galaxies and confirms our results with the much smaller samples. 

Just because AGN host galaxies have systematically lower sSFRs than non-AGN galaxies however does not directly imply that AGN feedback is taking place. Two scenarios can possibly explain our results. 1.) AGN actively quench star formation through short outbursts during the late life of the galaxy. 2.) AGN are simply the result of the availability of cold gas in a galaxy.

High-mass star-forming galaxies on the main sequence are gas rich especially in molecular gas \citep{Saintonge:2012uq}. This large cold gas supply can fuel both star-formation and AGN activity. If we prescribe to the ``bathtub'' model of gas regulation \citep{Lilly:2013pd}, the SFR is simply proportional to the mass of the gas reservoir in the galaxy. The two processes that regulate the mass are the accretion rate from the halo and a wind outflow, which in \citet{Lilly:2013pd} is proportional to the SFR. 

In the first scenario we can imagine the AGN significantly adding to the decrease of gas mass in the galaxy through several ways such as halo heating and powerful winds. All of these feedback processes work to reduce the mass of gas in the galaxy, which then reduces the SFR and produces the shift in sSFR we see compared to non-AGN star-forming galaxies (Figure~\ref{fig:dlogSFR_groups}). However based on Figure~\ref{fig:sSFR_prop} we know the SFR is not connected to the \textit{instantaneous} AGN strength, rather the SFR is correlated with the \textit{average} AGN strength over 100 Myr timescales \citep{Chen:2013uq}. So if AGN feedback is working it must be only over relatively short periods of time probably while undergoing a powerful outburst. This is supported by findings of molecular outflows in powerful AGN \citep{Veilleux:2013qq, Cicone:2014ty} where the mass outflow rate was shown to rise with increasing AGN luminosity. It is also supported by the discovery of ``voorwerpje'' \citep{Lintott:2009fk, Keel:2012qy, Keel:2014lr}, highly ionized clouds at kpc scales around currently dormant SMBH that indicate AGN outbursts on timescales of $\sim100,000$ years. \citet{Tombesi:2015fj} recently were able to show that a fast accretion-disk wind is driving a molecular outflow in a ULIRG. Within this framework, this means all of the BAT AGN lying below the MS have gone through at least one powerful phase or possibly more to be able to significantly deplete the galaxy of molecular gas. The AGN still inside the MS perhaps are still waiting for that outburst or haven't gone through enough to move off the MS.

In the second scenario, the existence of an AGN is the consequence of the large availability of cold gas in high mass galaxies. If there is enough cold gas, eventually enough will find its way to the centers to trigger an AGN. AGN would likely turn on while the galaxy is still on the MS. A quenching process unrelated to the AGN, possibly shock heating of the accreting halo gas, slowly shuts down star formation beginning in the outskirts of the galaxy. This would be supported by our findings that the BAT AGN are more compact in the FIR than normal star-forming galaxies \citep{Mushotzky:2014ad}. The AGN persists as the galaxy falls off the MS, eventually turning off as the remaining cold gas runs out. This scenario is still supported by the findings of \citet{Chen:2013uq}. The long timescale average accretion rate is tightly connected to the SFR through the available gas reservoir. It would only require that gas accretion onto the SMBH not be a smooth and constant process but more intermittent and variable which is supported by high resolution simulations  \citep{Hopkins:2010kx, Novak:2011ek}.

There is one component we are currently ignoring however: mergers and interactions. With 30 per cent of the BAT AGN involved in one, in both scenarios mergers could be the key ingredient for funneling cold gas to the nuclei of galaxies and igniting an AGN. \citet{Schawinski:2014cr} cites major mergers as the mechanism to cause fast quenching in early type galaxies whereas late type galaxies are slowly quenched through a drying up of their gas reservoir. Perhaps both quenching mechanisms are at play in the BAT AGN. The majority are contained in massive spirals (Figure~\ref{fig:morph_breakdown}) where AGN feedback (scenario 1) or external processes (scenario 2) are slowly suppressing star-formation while the ones involved in a merger have had their SFRs and SMBH accretion rates briefly elevated before rapidly falling off the main sequence. This could also explain why we still see a large fraction of mergers just below the main sequence. 

\begin{figure}
	\begin{centering}
		\includegraphics[width=\columnwidth]{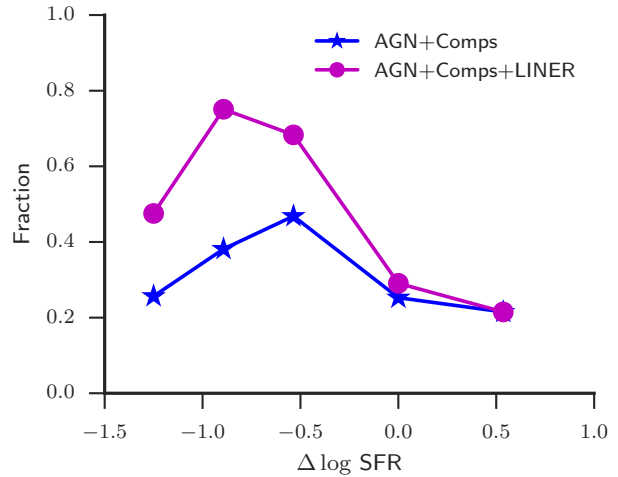}
		\caption{Fraction of all COLD GASS galaxies that are AGN or Composite galaxies (blue star) and AGN, Composite, or LINERs (magenta circles) as a function of distance from the main sequence. The fractions shown are the average fractions of all 50 representative subsamples. \label{fig:agn_fraction}}
	\end{centering}
\end{figure}

Which scenario is dominant is still a matter of debate and at present we are unable to distinguish between these concepts. One thing that is clear though is that massive galaxies go through an AGN phase as they fall off the main sequence. This is shown in Figure~\ref{fig:agn_fraction} where we plot the fraction of all galaxies that are classified as an AGN/Composite (blue line) or AGN/Composite/LINER (magenta line, if we suppose LINERs also contain an AGN) as a function of distance from the MS in the COLD GASS sample. The fractions are the average across all 50 representative subsamples. Both lines peak in the $1-2\sigma$ and $2-3\sigma$ below regions with close to 80 per cent of the galaxies in these regions containing an AGN or LINER. The question that remains is whether these AGN have had a substantial effect on the SFR in their host galaxy or they are just ``along for the ride'' off the main sequence. An answer could possibly come from surveys using integral field unit (IFU) spectroscopy such as MaNGA\footnote{\url{https://www.sdss3.org/future/manga.php}} and CALIFA \citep{Sanchez:2012fk} that provide spatially resolved spectra across the entire galaxy.

\section{Summary and Conclusions}
Using consistent measures of both stellar mass and SFR, we compared three samples of galaxies, one a set of local, non-AGN star-forming galaxies (HRS), one a complete set of AGN host galaxies selected from the \swift/BAT catalog, and one a mass-selected sample of both star-forming and AGN host galaxies (COLD GASS). From the HRS, we constructed our own ``main sequence'' relation and systematically analyzed the location of the samples with respect to the main sequence. Our main conclusions are as follows:
\begin{enumerate}
\item[1.] AGN host galaxies, both X-ray selected and optically selected, systematically lie below the main sequence, indicating reduced levels of star-formation. 

\item[2.] After splitting the AGN sample into regions of increased offset from the main sequence of star formation, we found no dependence of the offset on hard X-ray luminosity.  

\item[3.] Analysing the morphologies of the samples, we find that while the fraction of BAT AGN that are in spirals most closely resembles the star-forming sample, the sSFR's are more closely related to the optically selected AGN. The distribution of sSFR for spirals and ellipticals in the BAT AGN also best match the distribution for spirals and ellipticals for optically selected AGN. We find a significant increase in the merger fraction from 0.1 well below the main sequence up to 0.4 near, inside, and above it. The merger fractions for the BAT AGN are much higher than those for the COLD GASS AGN and star-forming galaxies.

\item[4.] These results are both model independent as well as unaffected by our selection criteria. A detailed comparison with \citet{Mullaney:2012gf} does not find any discrepancy between the two studies, and evolving our population of AGN to higher redshifts agrees well with their detection fractions. 

\item[5.] Expanding the analysis to the larger SDSS sample of galaxies shows the effect of AGN host galaxies lying below the main sequence also occurs in larger optically selected samples and confirms the previous findings of AGN preferentially occurring in galaxies in transition from star-forming to quiescence.

\item[6.] We discussed how the trend in sSFR can be explained by AGN feedback that reduces the supply of cold gas in the galaxy. This slowly suppresses star-formation through short periodic outbursts. It can also be explained if the occurrence of AGN is simply the result of the availability of cold gas. As star formation is quenched through other processes the AGN follows along and eventually fades as the cold gas runs out. In either case it is clear AGN are prevalent in massive galaxies currently falling off the main sequence.
\end{enumerate}

\section*{Acknowledgements}
The authors thank A. Saintonge for providing the merger classifications for COLD GASS as well as for thoughtful discussion. The authors are grateful to S. Veilleux for helpful discussions and a careful reading. We thank the anonymous referee for a detailed review that greatly improved the quality of this Paper. This research has made use of the NASA/IPAC Extragalactic Database (NED) which is operated by the Jet Propulsion Laboratory, California Institute of Technology, under contract with the National Aeronautics and Space Administration. This publication makes use of data products from the Wide-field Infrared Survey Explorer, which is a joint project of the University of California, Los Angeles, and the Jet Propulsion Laboratory/California Institute of Technology, funded by the National Aeronautics and Space Administration. Funding for SDSS-III has been provided by the Alfred P. Sloan Foundation, the Participating Institutions, the National Science Foundation, and the U.S. Department of Energy Office of Science. The SDSS-III web site is \url{http://www.sdss3.org/}. All figures in this publication were made using the MATPLOTLIB Python package \citep{Hunter:2007}. This research made use of Astropy, a community-developed core Python package for Astronomy \citep[][\url{http://www.astropy.org}]{Astropy:2013ek}.

\footnotesize{
\bibliographystyle{mn2e}
\bibliography{my_bib}
}

\appendix
\section{Example Fits to the SED of the BAT AGN and HRS}\label{sec:appendix}
In this Appendix, we provide to the reader example SED fits to both the BAT AGN and HRS as well as a brief description of the model and fitting procedure. We chose to fit our SEDs with the model described in \citet{Casey:2012jl}, which is a combination of an exponentially cutoff mid-infrared (MIR) power law and a single temperature greybody. The greybody has the standard form, $F_{\nu} \propto \nu^{\beta}B_{\nu}(T)$, where $\beta$ is the dust emissivity spectral index and $B_{\nu}(T)$ is the standard Planck function with temperature $T$, and has been shown to fit very well the FIR SEDs of both normal star-forming galaxies as a whole \citep{Gordon:2010qy, Galametz:2012uq, Auld:2013ge, Cortese:2014qq} and small star-forming regions within the galaxy \citep{Galametz:2012uq, Smith:2012fj}.

The MIR power law, with the form $F_{\lambda} \propto \lambda^{\alpha}e^{-(\lambda/\lambda_{\rm turn})^2}$, can be thought of as the sum of many hot dust components that combine to form an overall power law in the broadband SED \citep{Kovacs:2010zr}. \citet{Casey:2012jl} found that this simple model fits the observed SEDs of ultra luminous infrared galaxies (ULIRGs) very well compared to single greybodies and standard SED template libraries \citep[e.g.][]{Chary:2001lr}. While the physical heating mechanism is different between Seyfert galaxies and ULIRGS, the same power law shape is observed in MIR spectra of AGN \citep{Schweitzer:2006mz, Mullaney:2011yq} and motivates the addition of this component for the BAT AGN. The equation for the total model is then:

\begin{equation}\label{eqn:model}
F_{\nu} = N_{pwlw}\nu^{-\alpha}e^{-(\lambda/\lambda_{\rm turn})^2} + N_{grey}\nu^{\beta}B_{\nu}(T)
\end{equation}

\begin{figure*}
	\includegraphics[width=\textwidth]{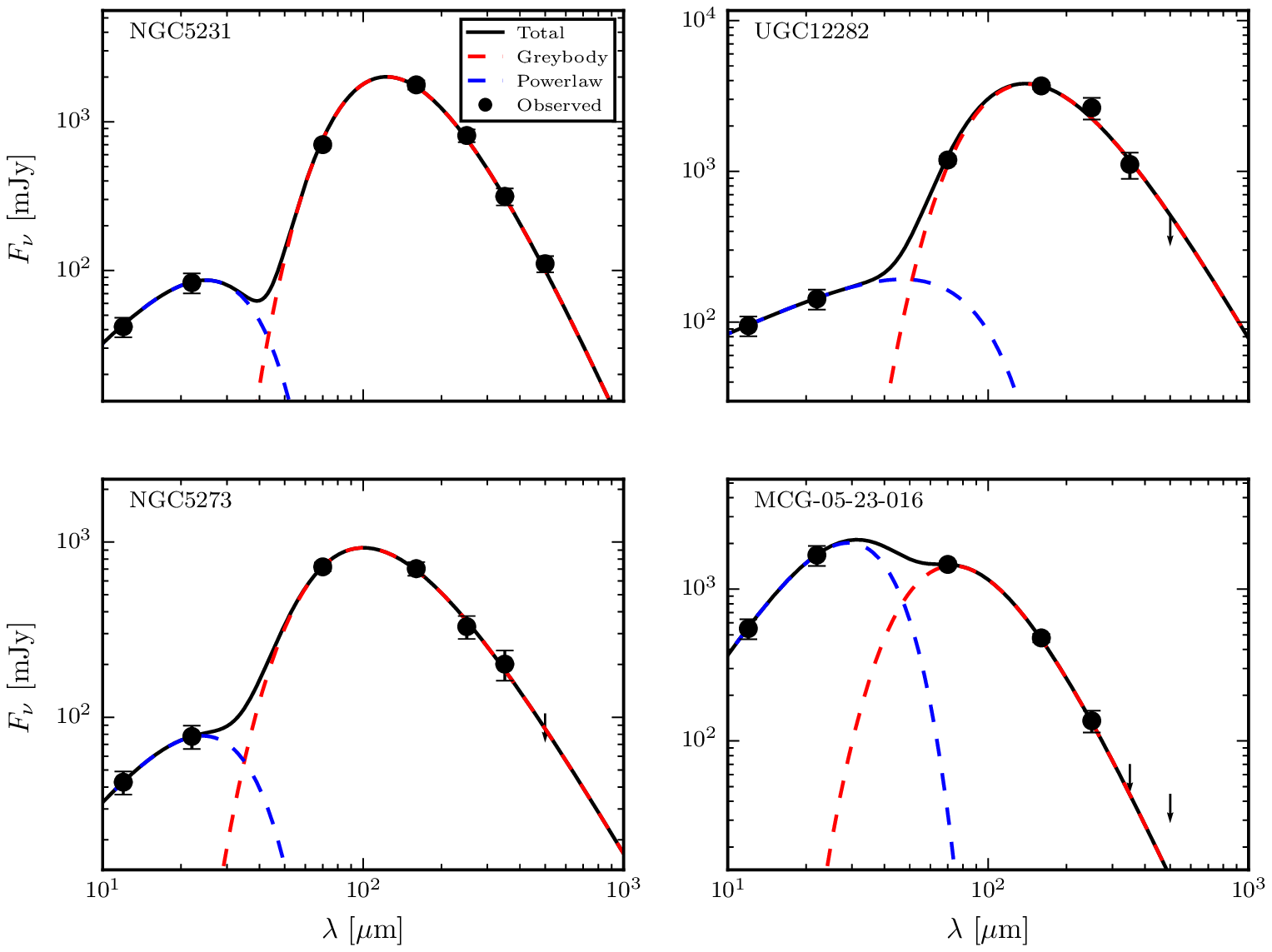}
	\caption{Example SED fits for four randomly chosen BAT AGN used for analysis in this Paper. The observed fluxes are shown as black dots with 1$\sigma$ error bars. 5$\sigma$ upper limits to the flux for undetected wavebands are shown as downward pointing arrows. The solid black line is the best fit model given by Equation~\ref{eqn:model} while the dashed blue and red lines indicate the best fit MIR powerlaw and greybody components, respectively.\label{fig:example_bat_sed}}
\end{figure*}

For each source we used all wavebands that had at least a 5$\sigma$ detection. The best fit parameters were found from $\chi^{2}$-minimization using the Levenberg-Marquardt algorithm for optimization. Figure~\ref{fig:example_bat_sed} displays four randomly chosen BAT AGN SEDs and their 1$\sigma$ uncertainties (black dots with error bars) along with the best fit model (black line). The dashed lines indicate the best fit MIR cutoff powerlaw (blue) and greybody (red). As shown, the model does quite well for the wide variety of SED shapes in the BAT AGN sample.

\begin{figure*}
	\includegraphics[width=\textwidth]{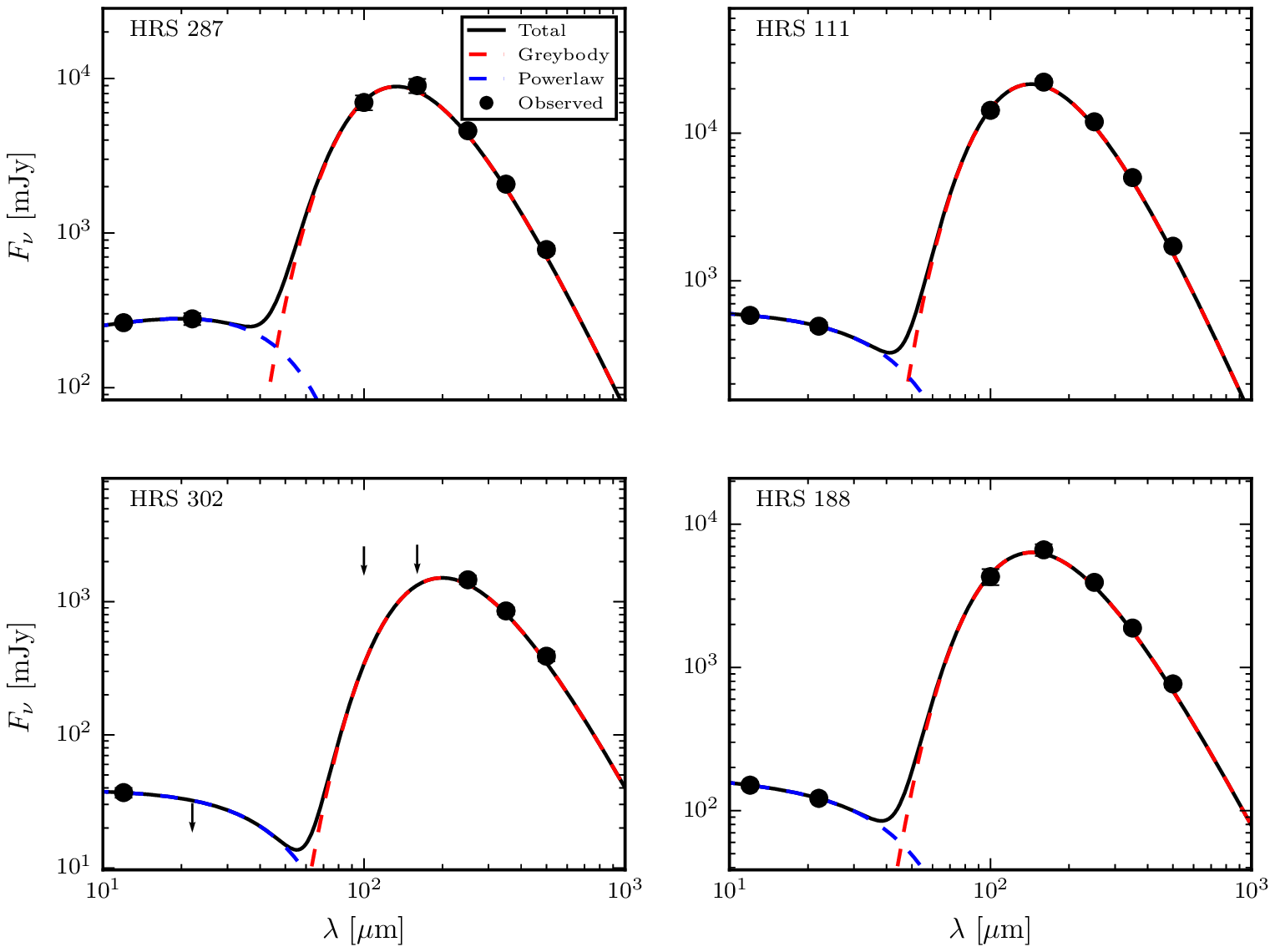}
	\caption{Same as Figure~\ref{fig:example_bat_sed} except for four randomly chosen HRS sources.\label{fig:example_hrs_sed}}
\end{figure*}

Figure~\ref{fig:example_hrs_sed} is the same as Figure~\ref{fig:example_bat_sed} except for four randomly chosen HRS sources to show the model works equally well for the sources without an AGN.

\end{document}